\documentclass[proceedings]{JHEP3} 

\PrHEP{ ahep2003}

\usepackage{graphicx}
\usepackage{epsfig,multicol}                    
\usepackage{cite}

\newbox\mybox
\newcommand\fverb{\setbox\mybox=\hbox\bgroup\verb}
\newcommand\fverbdo{\egroup\medskip\noindent\fbox{\unhbox\mybox}\ }
\newcommand\fverbit{\egroup\item[\fbox{\unhbox\mybox}]}

\newcommand {\ignore}[1]{}

\def\vb#1{\vbox to #1 pt{}}

\def\slash#1{#1\!\!\! /}

\def\ifmath#1{\relax\ifmmode #1\else $#1$\fi}
\def\half{\ifmath{{\textstyle{1 \over 2}}}}

\def\nn{\nonumber}

\def\ds{\displaystyle}
\def\beqa{\begin{eqnarray}}
\def\eeqa{\end{eqnarray}}
\def\ra{\rightarrow}

\def\Eq#1{Eq. (\ref{#1})}

\def\Fig#1{Fig.~\ref{#1}}

\def\te{{\tilde e}}
\def\tm{{\tilde \mu}}
\def\st{{\tilde \tau}}

\def\rp{$R_p \hspace{-1em}/\;\:$ }

\def\bold#1{\setbox0=\hbox{$#1$} 
     \kern-.025em\copy0\kern-\wd0 
     \kern.05em\copy0\kern-\wd0 
     \kern-.025em\raise.0433em\box0 }
\newcommand{\bemartin}[1]{\begin{equation} \label{(#1)}}
\newcommand{\eemartin}{\end{equation}}
\newcommand{\bamartin}[1]{\begin{eqnarray} \label{(#1)}}
\newcommand{\eamartin}{\end{eqnarray}} 

\newcommand {\chiz} [1] {\tilde{\chi}^{0}_{#1} }

\DeclareMathAlphabet{\mathsc}{OT1}{cmr}{m}{sc}
\newcommand{\Sol}  {\mathsc{sol}}
\newcommand{\Atm}  {\mathsc{atm}}
\newcommand{\Dms}  {\Delta m^2_\Sol}
\newcommand{\Dma}  {\Delta m^2_\Atm}
\newcommand{\eVq}  {\rm{eV}^2}

\title{Probing Neutrino Parameters at Accelerators}

\author{\speaker{Jorge C. Rom\~ao}\\
        Instituto Superior T\'ecnico, Departamento de F\'{\i}sica\\
        A. Rovisco Pais 1, 1049-001 Lisboa, Portugal\\
        E-mail: \email{jorge.romao@ist.utl.pt}
}

\conference{International Workshop on Astroparticle and High Energy Physics}

\abstract{The simplest unified extension of the Minimal Supersymmetric Standard
Model with bilinear R--Parity violation provides a predictive scheme
for neutrino masses and mixings
which can account for the observed atmospheric and
solar neutrino anomalies.
Despite the smallness of neutrino masses R-parity violation is
observable at present and future high-energy colliders, providing an
unambiguous cross-check of the model.}

\begin{document}

\section{Introduction}
The announcement of high statistics atmospheric neutrino data by the
\texttt{SuperKamiokande} collaboration \cite{fukuda:1998mi} has
confirmed the deficit of muon neutrinos, especially at small zenith
angles, opening a new era in neutrino physics.  Although in the past
have been considered alternative solutions for the atmospheric
neutrino anomaly ~\cite{maltoni:2002ni,gonzalez-garcia:2000sq} it is
now clear that the simplest interpretation of the data is in terms of
$\nu_{\mu}$ to $\nu_{\tau}$ flavor oscillations with maximal
mixing. This excludes a large mixing among $\nu_{\tau}$ and
$\nu_e$~\cite{fukuda:1998mi}, in agreement also with the
\texttt{CHOOZ} reactor data~\cite{apollonio:1999ae,boehm:2001ik}.  On
the other hand, the persistent disagreement between solar neutrino
data~\cite{fukuda:2002pe} and theoretical
expectations~\cite{bahcall:1998wm} has been a long-standing problem in
physics.  Recent solar neutrino data from the \texttt{Kamland}
collaboration~\cite{eguchi:2002dm} and the latest results from the
\texttt{SNO} collaboration~\cite{ahmed:2003kj}, clearly indicate that
we have MSW conversions with a large mixing angle
~\cite{maltoni:2003da}.  For the solar neutrino parameters we get for
the best fit point~\cite{maltoni:2003da}:
\begin{equation}
    \tan^2\theta_\Sol = 0.43, \qquad \Dms = 6.9\times10^{-5}~\eVq \, ,
\end{equation}
confirming that the solar neutrino mixing angle is large, but
significantly non-maximal. The $3\sigma$ region for $\theta$ is:
\begin{equation}
\label{thetasol.range}
    0.30 \leq \tan^2\theta_\Sol \leq 0.64,
\end{equation}
while the  $3\sigma$ region for $\Dms$ range is given by,
\begin{equation}
\label{sol.kam.ranges}
    5.4\times 10^{-5}~\eVq \leq \Dms \leq 9.5\times 10^{-5}~\eVq, 
\end{equation}
On the other hand, current atmospheric neutrino data require
oscillations involving $\nu_{\mu} \leftrightarrow \nu_{\tau}$
~\cite{fukuda:1998mi}. The most recent global analysis
gives~\cite{maltoni:2003da},
\begin{equation}
\label{t23d23}
    \sin^2\theta_\Atm = 0.52 \,,\: \Dma = 2.6 \times
    10^{-3}~\eVq \: 
\end{equation}
with the 3$\sigma$ ranges,
\begin{eqnarray}
\label{t23d23range} 
    0.31 \le \sin^2\theta_\Atm \le 0.72 \\ 
    1.4 \times 10^{-3}~\eVq \le \Dma  \le 3.7 \times
    10^{-3}~\eVq \,. 
\end{eqnarray}

Many attempts have appeared in the literature to explain the
data. Here we review recent
results~\cite{romao:1999up,hirsch:2000ef,porod:2000hv,
diaz:2003as,hirsch:2002ys,restrepo:2001me} obtained in a
model~\cite{diaz:1998xc} which is a simple extension of the Minimal
Supersymmetric Standard Model (\texttt{MSSM}) with bilinear R-parity
violation (\texttt{BRpV}).  This model, despite being a minimal
extension of the \texttt{MSSM}, can explain the solar and atmospheric
neutrino data. Its most attractive feature is that it gives definite
predictions for accelerator physics for the same range of parameters
that explain the neutrino data.

\section{The Model}

Since \texttt{BRpV} has been discussed in the literature several times
\cite{romao:1999up,hirsch:2000ef,diaz:1998xc,decampos:1995av,
akeroyd:1998iq,banks:1995by} we will repeat only the main features of
the model here.  We will follow the notation of
\cite{romao:1999up,hirsch:2000ef}.  The simplest bilinear \rp model
(we call it the \rp \texttt{MSSM}) is characterized by the
superpotential
\begin{equation}
\label{eq:Wpot} 
W = W_{MSSM} + W_{\slash R_P} 
\end{equation} 
In this equation $W_{MSSM}$ is the ordinary superpotential of the
\texttt{MSSM},
\begin{equation}
W=\varepsilon_{ab}\left[
 h_U^{ij}\widehat Q_i^a\widehat U_j\widehat H_u^b
+h_D^{ij}\widehat Q_i^b\widehat D_j\widehat H_d^a
+h_E^{ij}\widehat L_i^b\widehat R_j\widehat H_d^a 
-\mu\widehat H_d^a\widehat H_u^b \right]
\end{equation}
where $i,j=1,2,3$ are generation indices, $a,b=1,2$ are $SU(2)$
indices. We have three additional terms that break R-parity,
\begin{equation}
\label{eq:WRPV} 
W_{\slash R_P} = \epsilon_{ab} \epsilon_i \widehat
  L^a_i\widehat H^b_u.  
\end{equation} 
These bilinear terms, together with the corresponding terms in the
soft supersymmetric (\texttt{SUSY}) breaking part of the Lagrangian, 
\begin{equation}
\label{eq:Lsoft}
  V_{soft} = V_{soft}^{MSSM} +
  \epsilon_{ab} B_i \epsilon_i {\tilde L}^a_i H^b_u 
\end{equation} 
define the minimal model, which we will adopt throughout this paper.
The appearance of the lepton number violating terms in Eq.
(\ref{eq:WRPV}) leads in general to non-zero vacuum expectation values
for the scalar neutrinos $\langle {\tilde \nu}_i \rangle$, called
$v_i$ in the rest of this paper, in addition to the VEVs $v_U$ and
$v_D$ of the \texttt{MSSM} Higgs fields $H_u^0$ and $H_d^0$.  Together
with the bilinear parameters $\epsilon_i$ the $v_i$ induce mixing
between various particles which in the \texttt{MSSM} are distinguished
(only) by lepton number (or R--parity).  Mixing between the neutrinos
and the neutralinos of the \texttt{MSSM} generates a non-zero mass for
one specific linear superposition of the three neutrino flavor states
of the model at tree-level while 1-loop corrections provide mass for
the remaining two neutrino states~\cite{romao:1999up,hirsch:2000ef}.

\section{Neutrino Masses and Mixings}

\subsection{Tree Level Neutral Fermion Mass Matrix}

In the basis $\psi^{0T}=
(-i\lambda',-i\lambda^3,\widetilde{H}_d^1,\widetilde{H}_u^2, \nu_{e},
\nu_{\mu}, \nu_{\tau} )$ the neutral fer\-mions mass terms in the
Lagrangian are given by
\begin{equation}
\mathcal{L}_m=-\frac 12(\psi^0)^T{\bold M}_N\psi^0+h.c.   
\end{equation}
where the neutralino/neutrino mass matrix is 
\begin{equation}
{\bold M}_N=\left[  
\begin{array}{cc}  
\mathcal{M}_{\chi^0}& m^T \cr
m & 0 \cr
\end{array}
\right]
\end{equation}
with
\begin{equation}
\mathcal{M}_{\chi^0}=\left[  
\begin{array}{cccc}  
M_1 & 0 & -\frac 12g^{\prime }v_d & \frac 12g^{\prime }v_u \cr
0 & M_2 & \frac 12gv_d & -\frac 12gv_u \cr
-\frac 12g^{\prime }v_d & \frac 12gv_d & 0 & -\mu  \cr
\frac 12g^{\prime }v_u & -\frac 12gv_u &  -\mu & 0  \cr
\end{array}  
\right] 
\quad ; \quad
m=\left[  
\begin{array}{c}
a_1 \cr
a_2 \cr
a_3 
\end{array}  
\right] 
\end{equation}
where $a_i=(-\frac 12g^{\prime }v_i, \frac 12gv_i, 0,\epsilon_i)$. 
This neutralino/neutrino mass matrix is diagonalized by 
\begin{equation}
\mathcal{ N}^*{\bold M}_N\mathcal{N}^{-1}={\rm diag}(m_{\chi^0_1},m_{\chi^0_2}, 
m_{\chi^0_3},m_{\chi^0_4},m_{\nu_1},m_{\nu_2},m_{\nu_3}) 
\label{eq:NeuMdiag} 
\end{equation}

\subsection{Approximate Diagonalization at Tree Level }

If the \rp parameters are small (we will show below that this is
indeed the case), then we can block-diagonalize ${\bold M}_N$
approximately to the form diag($m_{eff},\mathcal{M}_{\chi^0}$)
\begin{equation}
m_{eff} = - m \cdot \mathcal{M}_{\chi^0}^{-1} m^T = 
\frac{M_1 g^2 + M_2 {g'}^2}{4\, \det(\mathcal{M}_{\chi^0})} 
\left(\begin{array}{ccc}
\Lambda_e^2 & \Lambda_e \Lambda_\mu
& \Lambda_e \Lambda_\tau \\
\Lambda_e \Lambda_\mu & \Lambda_\mu^2
& \Lambda_\mu \Lambda_\tau \\
\Lambda_e \Lambda_\tau & \Lambda_\mu \Lambda_\tau & \Lambda_\tau^2
\end{array}\right),
\end{equation}
with
\begin{equation}
\Lambda_i=\mu v_i + v_d \epsilon_i \, .
\end{equation}
The matrices $N$ and $V_{\nu}$ diagonalize 
$\mathcal{M}_{\chi^0}$ and $m_{eff}$ 
\begin{equation}
N^{*}\mathcal{M}_{\chi^0} N^{\dagger} = {\rm diag}(m_{\chi^0_i})
\quad ; \quad
V_{\nu}^T m_{eff} V_{\nu} = {\rm diag}(0,0,m_{\nu}),
\end{equation}
where 
\begin{equation}
m_{\nu} = Tr(m_{eff}) = 
\frac{M_1 g^2 + M_2 {g'}^2}{4\, \det(\mathcal{M}_{\chi^0})} 
|{\vec \Lambda}|^2.
\label{eq:mneu3}
\end{equation}

So we get at tree level only one massive neutrino, the two other
eigenstates remaining massless. The tree level value will give the
atmospheric mass scale, while the other states will get mass at one
loop level. Therefore the \texttt{BRpV} model produces a hierarchical mass 
spectrum. We will show below how the one loop masses are generated.

\subsection{One Loop Neutrino Masses and Mixings}

\subsubsection{Definition}

The Self--Energy for the neutralino/neutrino is~\cite{hirsch:2000ef},

\begin{equation}
\hskip 3.5cm \equiv
i \left\{ \slash{p} \left[ P_L \Sigma^L_{ij} + P_R \Sigma^R_{ij} \right]
-\left[ P_L \Pi^L_{ij} + P_R \Pi^R_{ij} \right]\right\}
\end{equation}

\begin{picture}(0,0)
\put(-0.5,-0.25){\includegraphics[width=3.5cm]{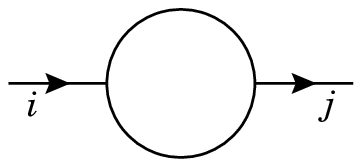}}
\end{picture}

\noindent
Then the pole mass is,
\begin{equation}
M^{\rm pole}_{ij}= M^{\rm \overline{DR}}_{ij}(\mu_R) + \Delta M_{ij}
\end{equation}
with
\begin{equation}
\Delta M_{ij}\! =\! \left[ \half 
\left(\Pi^V_{ij}(m_i^2)\! +\! \Pi^V_{ij}(m_j^2)\right) 
\!-\! \half 
\left( m_{\chi^0_i} \Sigma^V_{ij}(m_i^2) \! +\! 
m_{\chi^0_j} \Sigma^V_{ij}(m_j^2) \right) \right]_{\Delta=0}
\end{equation}
where
\begin{equation}
\Sigma^V=\half \left(\Sigma^L+\Sigma^R\right)
\quad ; \quad
\Pi^V=\half \left(\Pi^L+\Pi^R\right)
\end{equation}
an the parameter that appears in dimensional reduction is,
\begin{equation}
\ds \Delta=\frac{2}{4-d} -\gamma_E + \ln 4\pi
\end{equation}

\subsubsection{Diagrams Contributing}

In a generic way the diagrams contributing are given in
\Fig{fig:diags}, where all the particle in the model circulate in the
loops. 
\FIGURE
{
  \centering
  \begin{tabular}{ccc}
    \includegraphics[width=30mm]{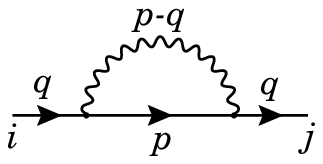}
    &
    \includegraphics[width=30mm]{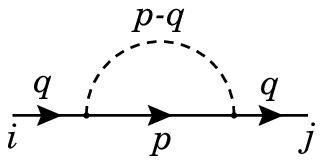}
    &
    \includegraphics[height=24mm]{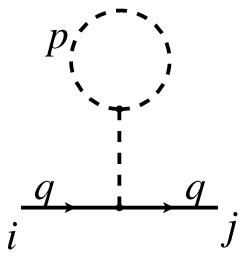}
  \end{tabular}
  \caption{Diagrams contributing at one loop to the neutrino mass matrix}
  \label{fig:diags}
}
These diagrams can be calculated in a straightforward way. For
instance the $W$ diagram in the $\xi=1$ gauge gives
\begin{eqnarray}
\Sigma^V_{ij}&\!=\!\!& -\frac{1}{16\pi^2}\, \sum_{k=1}^5
2 \left(O^{\rm ncw}_{L jk} O^{\rm cnw}_{L ki} +
O^{\rm ncw}_{R jk} O^{\rm cnw}_{R ki}\right) B_1(p^2,m^2_k,m^2_W)\cr
\vb{35}
\Pi^V_{ij}&\!=\!\!& -\frac{1}{16\pi^2}\, \sum_{k=1}^5
(-4) \left(O^{\rm ncw}_{L jk} O^{\rm cnw}_{R ki} +
O^{\rm ncw}_{R jk} O^{\rm cnw}_{L ki}\right) m_k\, B_0(p^2,m^2_k,m^2_W)
\end{eqnarray}
where $B_0$ and $B_1$ are the Passarino-Veltman
functions~\cite{passarino:1979jh},  and
$O^{\rm cnw}$, $O^{\rm ncw}$ are coupling matrices. Explicit
expressions can be found in~\cite{hirsch:2000ef}.

\subsubsection{Gauge Invariance}

When calculating the self--energies the question of gauge invariance
arises. We have performed a careful calculation in an arbitrary
$R_{\xi}$ gauge and showed~\cite{hirsch:2000ef} 
that the result was independent of the
gauge parameter $\xi$.

\section{Results for the Solar and Atmospheric Neutrinos}
\label{sec:neutrino-results}

\subsection{The masses}

\FIGURE[ht]{
  \includegraphics[width=80mm]{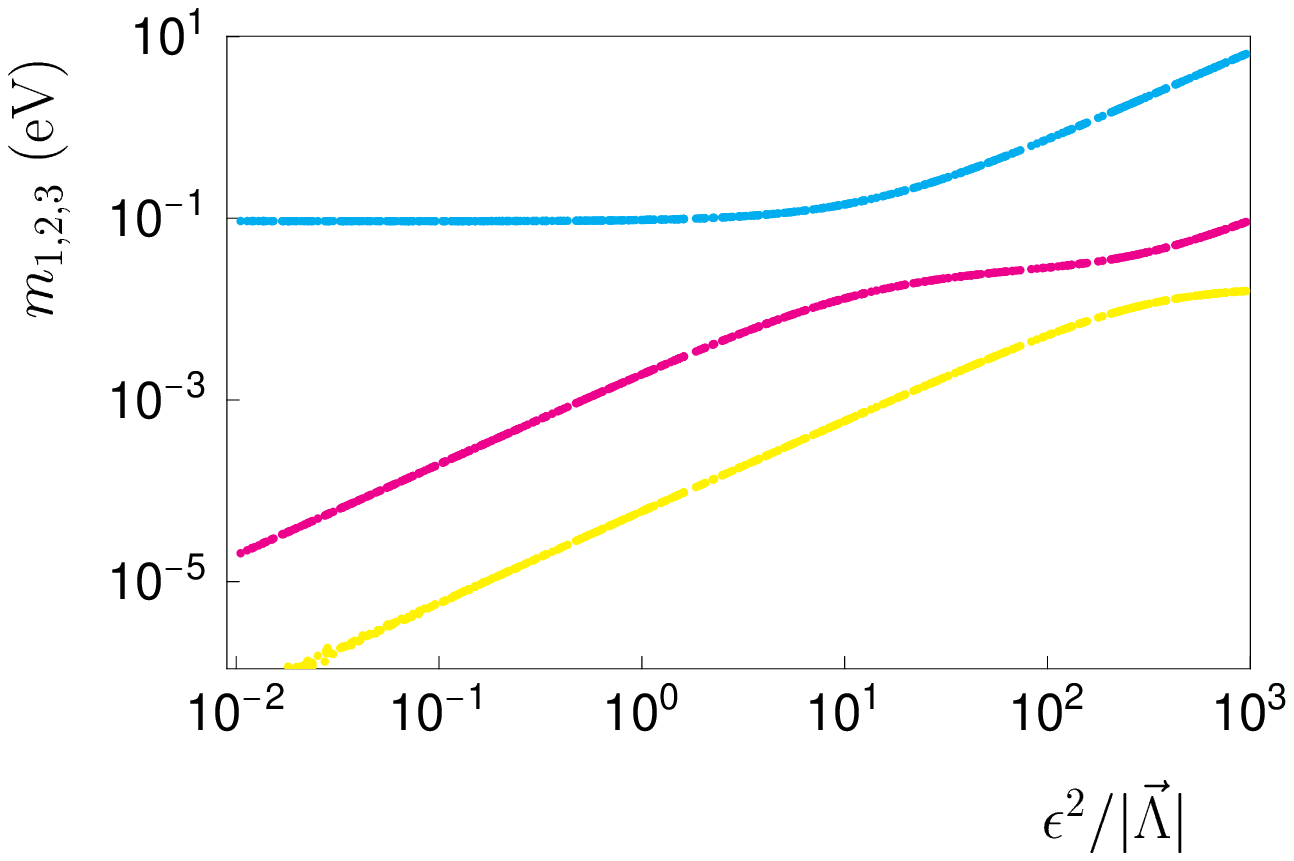}
  \vspace{-5mm}  
  \caption{neutrino masses}
  \label{fig:numasses}
}

The \texttt{BRpV} model produces a hierarchical mass spectrum for
almost all choices of parameters. The largest mass can be estimated by
the tree level value using Eq.~(\ref{eq:mneu3}).  Correct $\Delta
m^2_{\Atm}$ can be easily obtained by an appropriate choice of $| \vec
\Lambda|$. The mass scale for the solar neutrinos is generated at
1--loop level and therefore depends in a complicated way in the model
parameters. We will see below how to get an approximate formula for
the solar mass valid for most cases of interest.  Here we just present
in Fig.~\ref{fig:numasses}, for illustration purposes, the plot of the
three eigenstates as a function of the parameter
$\epsilon^2/|\Lambda|$, for a particular values of the \texttt{SUSY}
parameters, $m_0=\mu=500$ GeV, $\tan \beta= 5$, $B=-A=m_0$. For the
R-parity parameters we took $|\vec \Lambda|=0.16$ GeV, $10*
\Lambda_e=\Lambda_{\mu}=\Lambda_{\tau}$ and
$\epsilon_1=\epsilon_2=\epsilon_3$.

\subsection{The mixings}

Now we turn to the discussion of the mixing angles.  As can be seen
from \Fig{fig:numasses}, if $\epsilon^2/|\vec \Lambda| \ll 100$, then
the 1--loop corrections are not larger than the tree level results. In
this case the flavor composition of the 3rd mass eigenstate is
approximately given by
\begin{equation}
U_{\alpha 3}\approx\Lambda_{\alpha}/|\vec \Lambda |
\end{equation}
As the atmospheric and reactor neutrino data tell us that
$\nu_{\mu}\ra \nu_{\tau}$ oscillations are preferred over 
$\nu_{\mu}\ra \nu_e$, we conclude that  
\begin{equation}
\Lambda_e \ll \Lambda_{\mu} \simeq \Lambda_{\tau}
\end{equation}
are required for \texttt{BRpV} to fit the data. This is sown in
Fig.~\ref{corfu_fig2} a). We cannot set all the $\Lambda_i$ equal,
because in this case $U_{e 3}$ would be too large contradicting the
\texttt{CHOOZ} result as shown in Fig.~\ref{corfu_fig2} b).
\FIGURE{
\begin{tabular}{cc}
\includegraphics[width=0.45\textwidth]{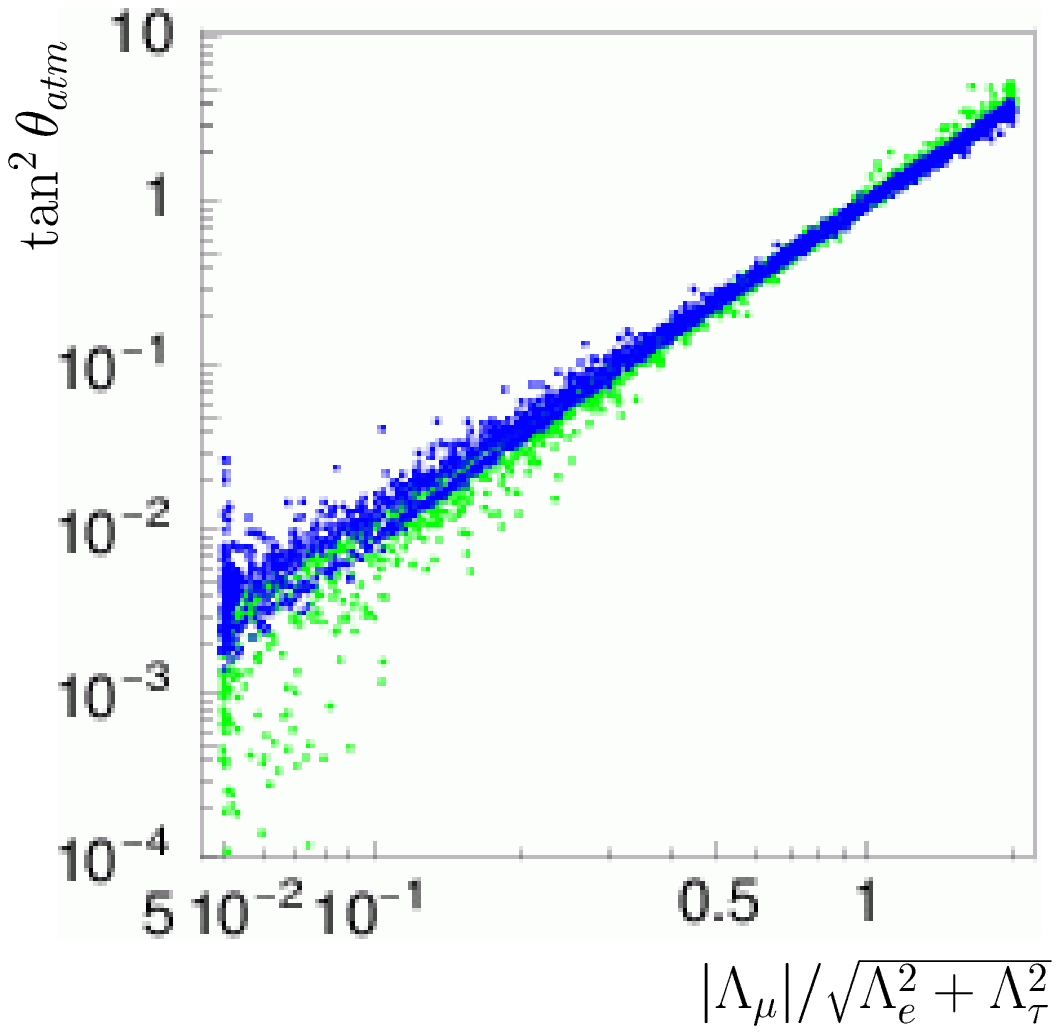}
&
\includegraphics[width=0.45\textwidth]{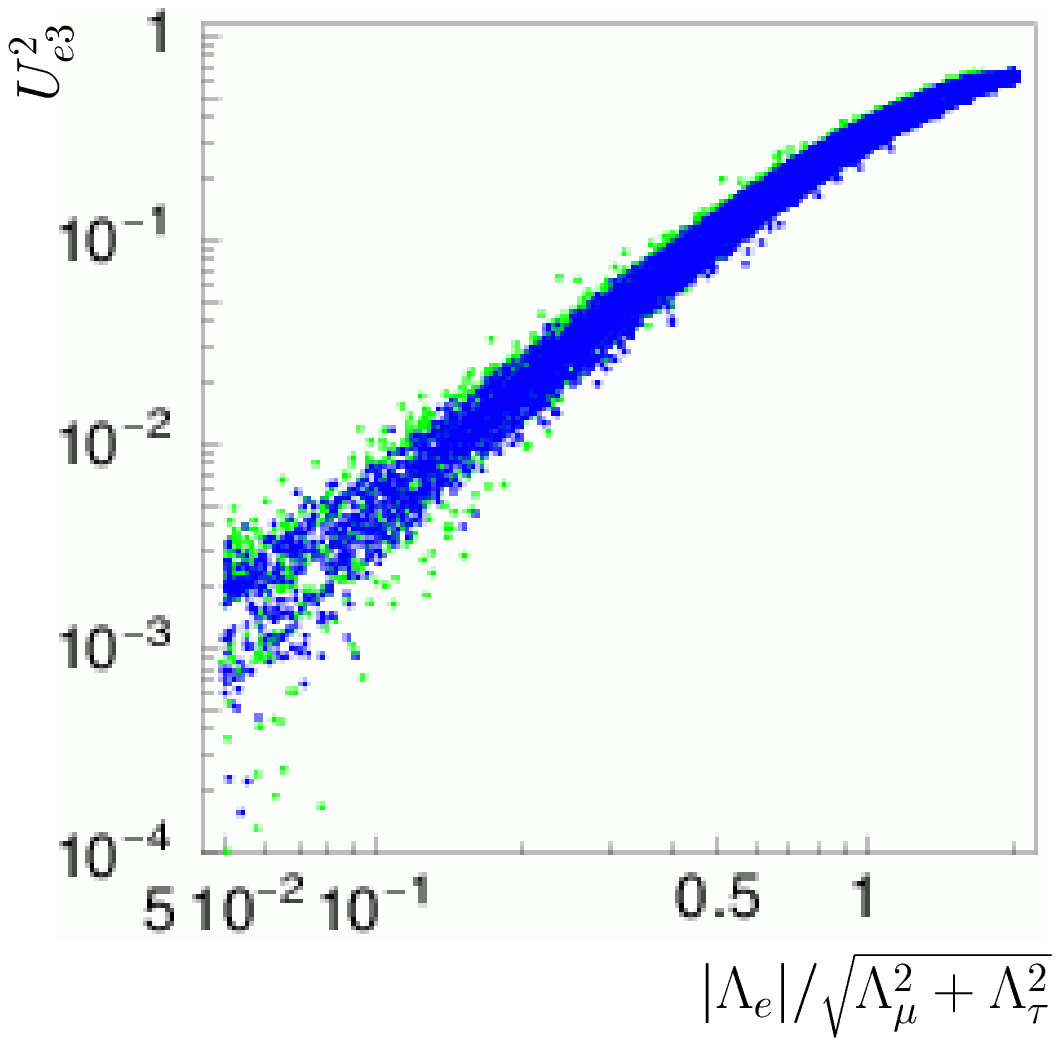}
\end{tabular}
\vspace{-2mm}
\caption{\small 
a) Atmospheric angle as a function of
  $|\Lambda_{\mu}|/\sqrt{\Lambda^2_{e}+\Lambda^2_{\tau}}$. 
  b) $U^2_{e3}$ as a function of
  $|\Lambda_e|/\sqrt{\Lambda^2_{\mu}+\Lambda^2_{\tau}}$.  
}
\label{corfu_fig2}
}
We have then two scenarios. In the first one, that we call the
\texttt{mSUGRA} case, we have universal boundary conditions of the
soft \texttt{SUSY} breaking terms. In this case we can
show~\cite{romao:1999up,hirsch:2000ef} that
\begin{equation}
  \label{eq:msugra}
\frac{\epsilon_e}{\epsilon_{\mu}}\simeq \frac{\Lambda_e}{\Lambda_{\mu}}  
\end{equation}
Then from Fig.~\ref{corfu_fig2} b) and the \texttt{CHOOZ} constraint
on $U^2_{e3}$, we obtain that \textit{both} ratios in
Eq.~(\ref{eq:msugra}) have to be small. Then from
Fig.~\ref{corfu_fig3} we conclude that the only possibility is the
small angle mixing solution for the solar neutrino problem. In the
second scenario, which we call the \texttt{MSSM} case, we consider
non--universal boundary conditions of the soft \texttt{SUSY} breaking
terms. We have shown that even a very small deviation from
universality (less then 1\%) of the soft parameters at the
\texttt{GUT} scale relaxes this constraint.  In this case
\begin{equation}
\frac{\epsilon_e}{\epsilon_{\mu}}\not=\frac{\Lambda_e}{\Lambda_{\mu}}  
\end{equation}
Then we can have at the same time \textbf{small} $U_{e3}^2$ determined
by $\Lambda_e/\Lambda_{\mu}$ as in Fig.~\ref{corfu_fig2} b) and
\textbf{large} $\tan^2(\theta_{\Sol})$ determined by
$\epsilon_e/\epsilon_{\mu}$ as in Fig.~\ref{corfu_fig3} b). After the
\texttt{Kamland} and \texttt{SNO} salt results, this is the only scenario
consistent with the data.

\FIGURE{
\begin{tabular}{cc}
\includegraphics[width=0.45\textwidth,height=60mm]{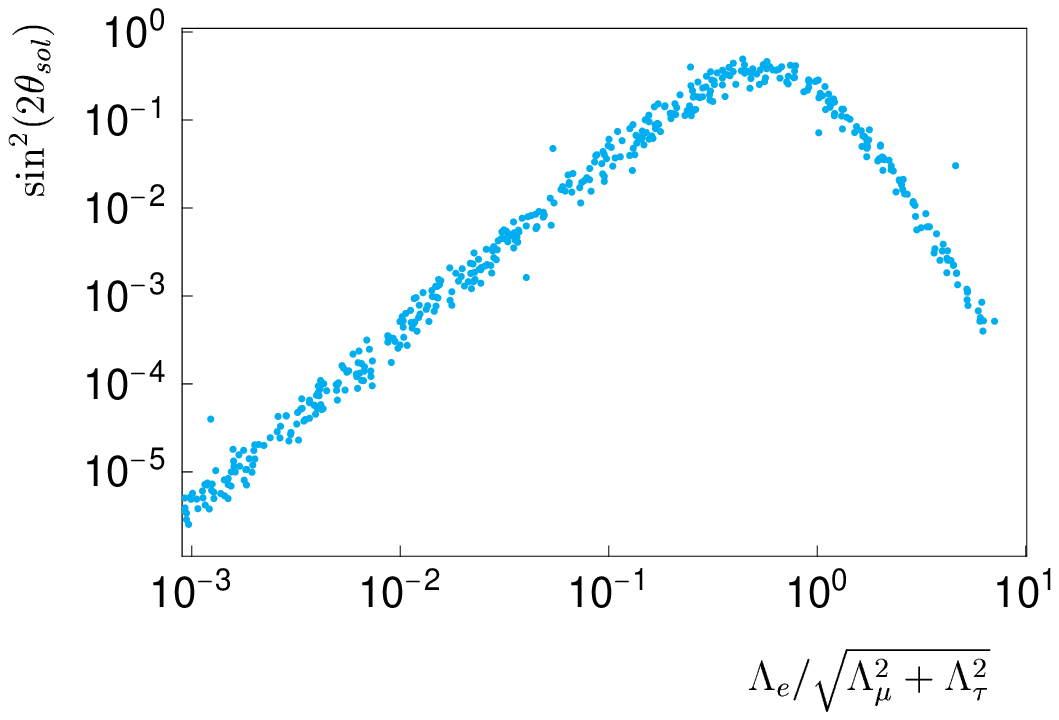}
\vspace{-5mm}
&
\includegraphics[width=0.44\textwidth,height=59mm,bb=113 479 413 799]{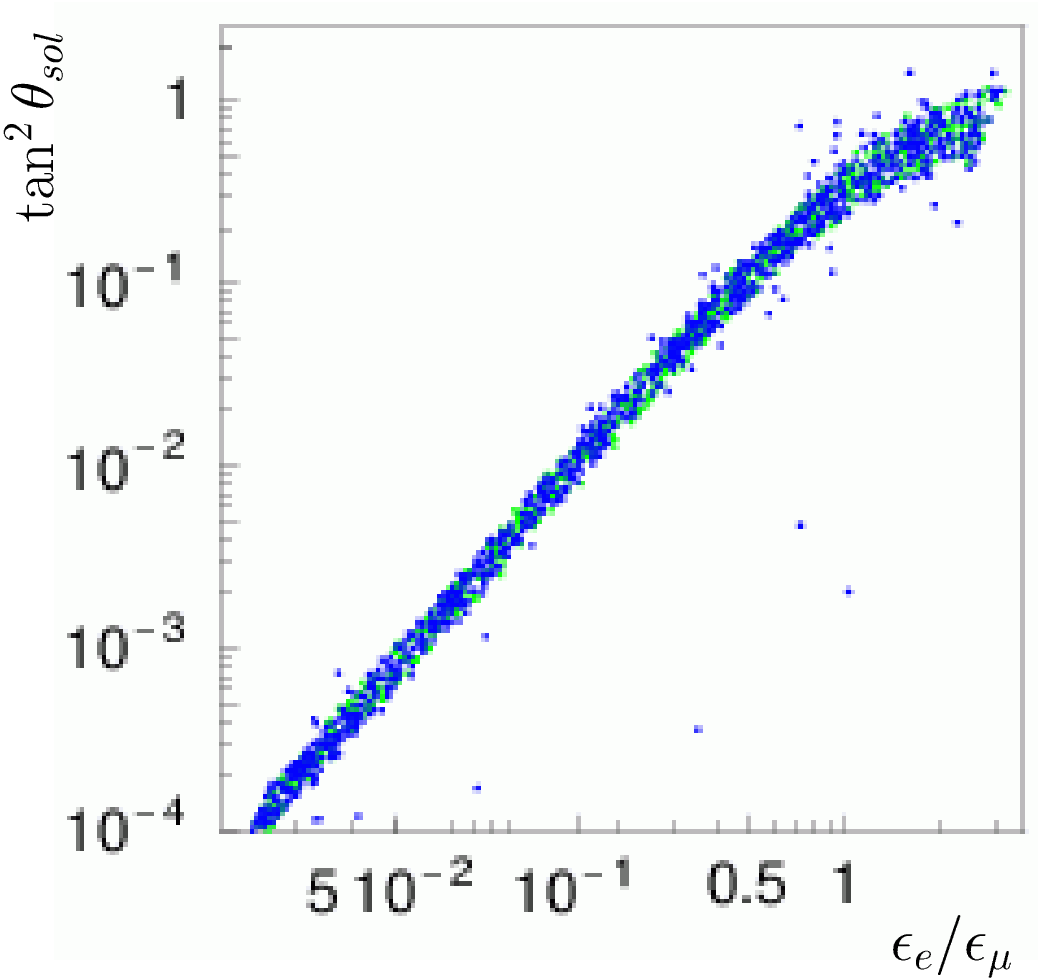} 
\end{tabular}
\caption{\small
Solar angle as function of: a)
$|\Lambda_e|/\sqrt{\Lambda^2_{\mu}+\Lambda^2_{\tau}}$ \ ;  
b) $\epsilon_e/\epsilon_{\mu}$.
}
\label{corfu_fig3}
}

\section{Approximate formulas for the solar mass and mixing}

In all the previous analysis we used a numerical program to evaluate
the one-loop masses and mixings. It is however desirable to have
analytical approximate results that can give us quickly the most
important contributions.  We have identified that these are the 
bottom-sbottom loops and the charged scalar-charged fermion loops.
Then, by expanding in powers of the small R-Parity breaking parameters
$\epsilon_i$, we get approximate formulas for the solar mass scale and 
mixing angle as we will explain below.

\subsection{Bottom-sbottom loops}

The contribution from the bottom-sbottom loop can be expressed
as\cite{diaz:2003as},

\begin{equation}
\Delta M_{ij}=-{{N_c m_b}\over{16\pi^2}}
2s_{\tilde b}c_{\tilde b}h_b^2
\Delta B_0^{\tilde b_1\tilde b_2}
\left[
\displaystyle 
\frac{\tilde\epsilon_i\tilde\epsilon_j}{\mu^2}
  + a_3 b \left(\tilde\epsilon_i\delta_{j3}+
 \tilde\epsilon_j\delta_{i3}\right)|\vec\Lambda| 
 + \left( a_3^2 + \frac{ a_L a_R}{h^2_b}\right)
 \delta_{i3}\delta_{j3} 
 |\vec\Lambda|^2 \right] \, ,
\label{eq:bot-sbot}
 \end{equation}
where $\widetilde \epsilon_i$ are the $\epsilon_i$ in the basis where
the tree level neutrino mass matrix is diagonal,
\begin{equation}
\widetilde \epsilon_i = 
\left(V_\nu^{(0)T}\right)^{ij} \epsilon_j \, ,
\end{equation}
the $a_i$ are functions of the \texttt{SUSY} parameters, and
\begin{equation}
  \label{eq:3}
 \Delta B_0^{\tilde b_1\tilde b_2}=
  B_0(0,m_b^2,m_{\tilde b_1}^2)-B_0(0,m_b^2,m_{\tilde b_2}^2)\, .
  \end{equation}
The different contributions can be understood as coming from different
types of insertions as shown in \Fig{fig:botsbot}. In this figure 
open circles correspond to small R-parity violating projections, 
full circles to R-parity conserving projections, and
open circles with a cross inside to mass insertions which flip
chirality. With this understanding one can make a one to one
correspondence between Eq.~(\ref{eq:bot-sbot}) and \Fig{fig:botsbot}.

\FIGURE{
  \begin{tabular}{cc}
    \includegraphics[width=0.45\linewidth]{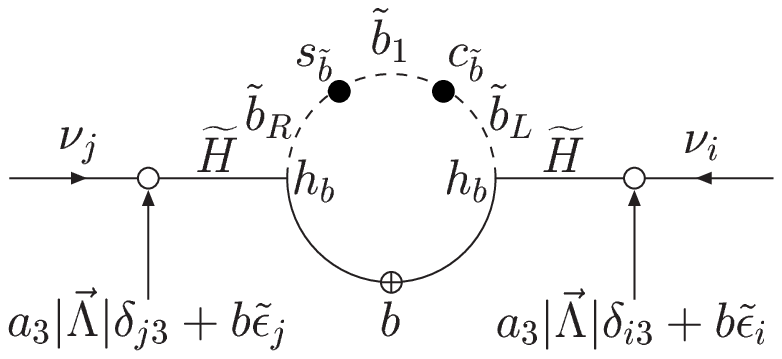}
    &\includegraphics[width=0.45\linewidth]{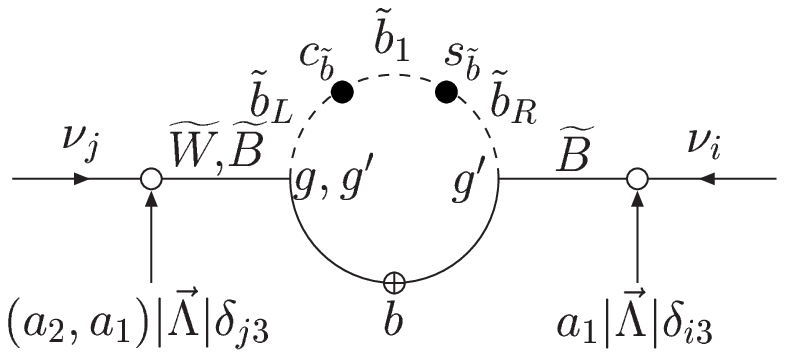}
  \end{tabular}
  \caption{Bottom-sbottom loop and different types of insertions.}
  \label{fig:botsbot}
}

\subsection{Charged Scalar-Charged Fermion Loops}

 Here the situation is much more complicated. As the charged scalars
 are now combinations of the six sleptons and of the two \texttt{MSSM}
 charged Higgs bosons we have many more particles that can be
 exchanged. For instance, the contribution from the staus is given in
 \Fig{fig:chscalar}. We see that, even for the staus, we have more
 diagrams than for the case of the bottom-sbottom loop. This is due
 to the fact that the breaking of R-parity is in the leptonic sector
 of the theory. So we can have R-parity violating insertions both on
 the external neutrino legs as well as in the particles circulating in
 the loop. A complete description of all the terms contributing to
 this loop can be found in Ref.\cite{diaz:2003as}, where one can also
 find the analytical expressions.

\FIGURE{
   \begin{tabular}{cc} 
     \includegraphics[width=0.45\linewidth]{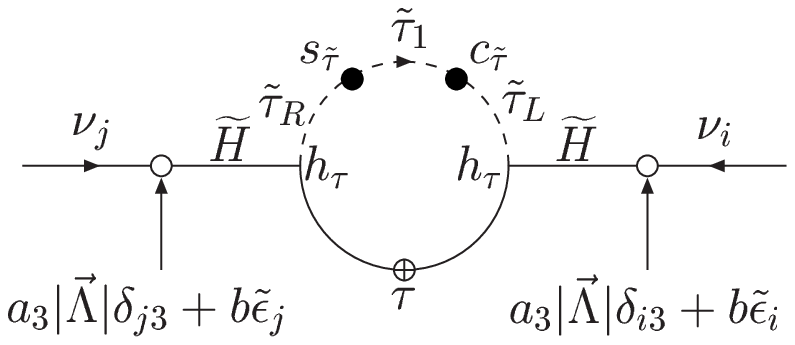}
     &\includegraphics[width=0.45\linewidth]{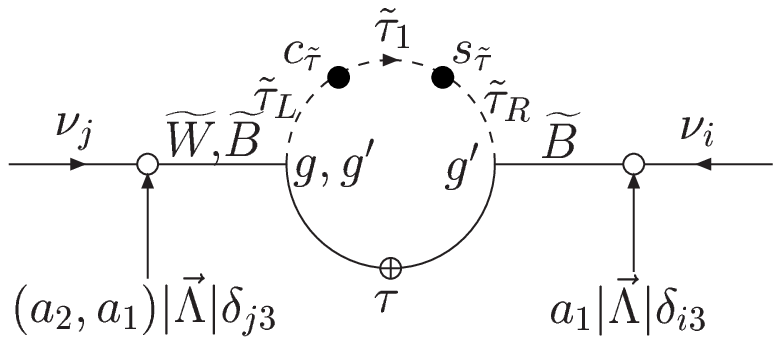}\\
     \includegraphics[width=0.45\linewidth]{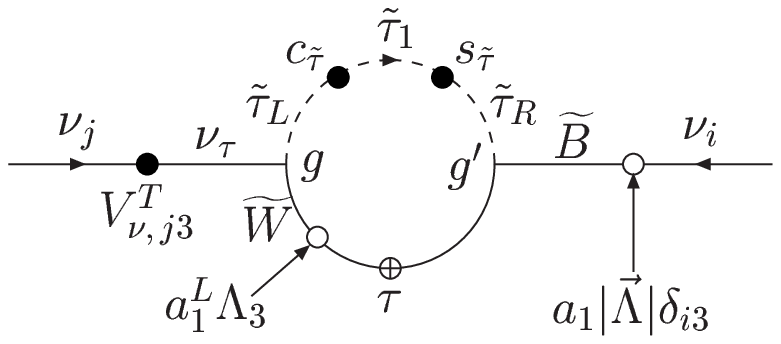}
     &\includegraphics[width=0.45\linewidth]{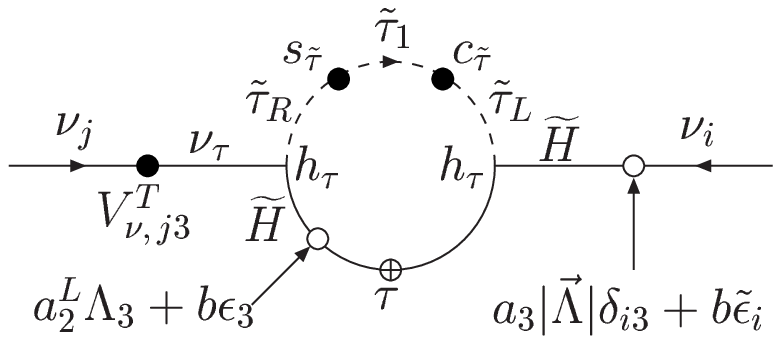}
   \end{tabular}
   \caption{Charged scalar loops: the stau contribution.}
   \label{fig:chscalar}
}

\subsection{Analytical vs Numerical results}

We now compare the approximate analytical formulas with the full
numerical calculation. This is shown in Fig.~\ref{fig:anvsnum}. On the
left panel we consider a data sample where the neutralino is the
Lightest Supersymmetric Particle (\texttt{LSP}). 
Shown are also the bands for LOW and LMA solutions of
the solar neutrino problem, as well as the much narrower band obtained
after the recent \texttt{Kamland}~\cite{eguchi:2002dm} and
\texttt{SNO} salt~\cite{ahmed:2003kj} results. On the right panel we
have the same situation for a data sample where the scalar tau is the
\texttt{LSP}. We see that the agreement is specially good (better than
10\%), just in the region which is compatible with the most recent
data.

\FIGURE{
  \label{fig:anvsnum}
  \begin{tabular}{cc}
    \includegraphics[width=0.45\linewidth]{NfDelM-v5.eps}
    &\includegraphics[width=0.45\linewidth]{Stau4aDelM-v5.eps}
  \end{tabular}
  \caption{Comparison between the analytical and numerical results for
    the solar mass. In
    the left panel the \texttt{LSP}=$\chi^0$ while in the right panel 
    the \texttt{LSP}=$\tilde{\tau}$. The red (dark) 
    band corresponds to the latest
    neutrino data~\cite{maltoni:2003da}.}
}

\subsection{Simplified approximation formulas}

In the basis where the tree-level neutrino mass matrix is diagonal the
mass matrix at one--loop level can be written as 
\vspace{-1mm}
\begin{equation}
  \widetilde m_\nu= V_\nu^{(0)T} m_\nu V_\nu^{(0)} 
      =\left( \begin{array}{ccc}
      c_1 \widetilde \epsilon_1 \widetilde \epsilon_1 & c_1 \widetilde
      \epsilon_1 \widetilde \epsilon_2  
      & c_1 \widetilde \epsilon_1 \widetilde \epsilon_3 \nonumber\\[+2mm]
      c_1 \widetilde \epsilon_2 \widetilde \epsilon_1 & c_1 \widetilde
      \epsilon_2 \widetilde \epsilon_2  
      & c_1 \widetilde \epsilon_2 \widetilde \epsilon_3 \nonumber\\
      c_1 \widetilde \epsilon_3 \widetilde \epsilon_1 & c_1 \widetilde
      \epsilon_3 \widetilde \epsilon_2  
      & c_0 |\vec\Lambda|^2 
       + c_1 \widetilde \epsilon_3 \widetilde \epsilon_3 
     \end{array} \right) + \cdots 
\label{eq:approx-mat}
\end{equation}
where
\begin{eqnarray}
c_0&=& \frac{M_1 g^2 + M_2 {g'}^2}{4\, \textrm{det}({\cal M}_{\chi^0})} 
 \\[+2mm]
c_1
&=& \frac{3}{16 \pi^2}\, \sin(2\theta_{\tilde b})\, 
m_b\, \Delta B_0^{\tilde b_2\tilde b_1}\ 
\frac{1}{\mu^2}
\end{eqnarray}
The dots in Eq.~(\ref{eq:approx-mat}) correspond to the terms that are
not proportional to the $\tilde \epsilon_i \times \tilde \epsilon_j$
structure, as can be seen from Eq.~(\ref{eq:bot-sbot}). 
Assuming that the bottom-sbottom loop dominates, the $\tilde
\epsilon_i \times \tilde \epsilon_j$ structure is dominant, and the
matrix can be diagonalized approximately under the condition
\begin{equation}
x\equiv\frac{c_1 |\vec{\widetilde{\epsilon}}|^2}{c_0 |\vec \Lambda|^2 }\ll 1
\end{equation}
Then we get

\begin{equation}
  \label{eq:1}
  m_{\nu_2} \simeq \frac{3}{16 \pi^2} \sin(2\theta_{\tilde b}) 
  m_b \Delta B_0^{\tilde b_2\tilde b_1}\ 
  \frac{({\tilde \epsilon}_1^2 + {\tilde \epsilon}_2^2)}{\mu^2}
\end{equation}
and
\begin{equation}
  \label{eq:2}
  \tan^2\theta_\Sol  = \frac{\widetilde  \epsilon_1^2}{\widetilde
    \epsilon_2^2} 
\end{equation}

\smallskip

The results for the masses are presented in Fig.~\ref{fig:xx3}. 
\FIGURE{
  \begin{tabular}{cc}
    \includegraphics[width=0.45\linewidth]{Nf2002Sim-v5.eps}
    &\includegraphics[width=0.45\linewidth]{StauSim-v5.eps}
  \end{tabular}
  \caption{Comparison between the simplified analytical formulas 
    and numerical results for the solar mass. In
    the left panel the \texttt{LSP}=$\chi^0$ while in the right panel 
    the \texttt{LSP}=$\tilde{\tau}$. The red (dark) band 
    corresponds to the latest
    neutrino data~\cite{maltoni:2003da}.}
  \label{fig:xx3}
}
As in Fig.~\ref{fig:anvsnum}, on the left panel we have the data set where the
neutralino is the \texttt{LSP} while on the right panel the \texttt{LSP} is the stau.
We see that the agreement is fairly good, particularly in the region
allowed by the present data.
\FIGURE{
  \begin{tabular}{cc}
    \includegraphics[width=0.45\linewidth]{Nfnd0Angle-v4.eps}
    &\includegraphics[width=0.45\linewidth]{Nfnd1Angle-v4.eps}
  \end{tabular}
  \caption{Comparison between the simplified analytical formulas 
    and numerical results for the solar angle. In
    the left panel we present all points, and show in green the band
    that corresponds to more than 90\% of the points. In the right
    panel we show all the points that have $\sin(2\theta_{\tilde b}) \Delta
    B_0^{\tilde\tau_2\tilde\tau_1}> 0.02$. The red (dark) band corresponds to
    the latest neutrino data~\cite{maltoni:2003da}.}
  \label{fig:xx4}
}
For the solar angle the results are shown in \Fig{fig:xx4}. We see
that the agreement is not as good as for the masses, even if we
restrict ourselves to the present allowed values of
$\tan^2\theta_{\hbox{\scriptsize SOL}}$, shown by the red (dark) band (it
corresponds to $3\sigma$ errors as taken from~\cite{maltoni:2003da})
on the figure. However, as we can see on the left panel, for more of
90\% of the points the agreement is within 20\%. This corresponds to
the cases where the bottom-sbottom loop dominates as can be seen on
the right panel, where we applied the cut $\sin(2\theta_{\tilde b})
\Delta B_0^{\tilde\tau_2\tilde\tau_1}> 0.02$ to ensure that the
bottom-sbottom loop is not negligible.

So, in summary, we can say that the simplified approximate formulas of
Eq.~(\ref{eq:1}) and Eq.~(\ref{eq:2}) give a very good approximation
of the full result for most of the cases. Only in the case when the
bottom-sbottom loop is not the dominant diagram (small mixing) we get
large deviations.

\section{Rare radiative lepton decays}

 As the parameters involved in the R--parity violating operator are
constrained in order to predict neutrino masses in the sub-eV range,
we have addressed~\cite{carvalho:2002bq} the question of whether this
operator will induce rates for charged LFV processes of experimental
interest. Some of them occur at tree--level such as double $\beta$
decay \cite{faessler:1998db,hirsch:1998kc} and $\mu-e$ conversion in
nuclei \cite{faessler:2000pn}. One loop LFV decays as $l_j \rightarrow
l_i \gamma$ become interesting on this framework due to the
experimental interest in improving the current limits
\cite{groom:2000in}:
\begin{eqnarray}
BR(\mu \to e \gamma) &<& 1.2 \times 10^{-11} \nn\\ BR(\tau \to \mu
\gamma) &<& 1.1 \times 10^{-6} \nn\\ BR(\tau \to e \gamma) &<& 2.7
\times 10^{-6}.
\label{limit}
\end{eqnarray}
We have shown that the predictions for the last two processes are much
lower than the above limits and will not constrain the \texttt{BRpV}
model.  For $\mu \rightarrow e \gamma$ the predictions are compatible
with the current limit but could begin to constrain the model for the
bounds that will be reached in current \cite{Yashima:2000qz} or
planned experiments \cite{kuno:prism}, if only the atmospheric
neutrino data were taken in account.  However, the requirement that
the one--loop induced $\Delta m^2_{\Sol}$ is in agreement with the solar
neutrino data, implies that the predicted rates for $\mu
\rightarrow e \gamma$ will not be visible~\cite{carvalho:2002bq}, even
in those new experiments. This can be seen in \Fig{fig:mueg}, where it is
shown the contour plot for BR($ \mu \rightarrow e\gamma$) as well as
the maximum values of $\epsilon_1,\epsilon_2$ compatible with the
solar neutrino mass. So, in conclusion, the experimental constraints
on these rare decays do not constrain the \texttt{BRpV} model.
\FIGURE{
 \includegraphics[width=100mm]{areaa1.eps}
 \caption{Contour plots for BR($ \mu \rightarrow e\gamma$). The dashed
 line corresponds to $m_{\nu_2}=0.01$~eV.} 
 \label{fig:mueg}
}

\section{Probing Neutrino Mixing via \texttt{SUSY} Decays}

After having shown, in the previous sections, that the \texttt{BRpV}
model produces an hierarchical mass spectrum for the neutrinos that
can accommodate the present data for neutrino masses and mixings, we
now turn to accelerator physics and will show how the neutrino
properties can be probed by looking at the decays of supersymmetric
particles.


If R-parity is broken the \texttt{LSP} will decay. If the \texttt{LSP}
decays then cosmological and astrophysical constraints on its nature
no longer apply. Thus, within R-parity violating \texttt{SUSY}, a
priori {\em any} superparticle could be the \texttt{LSP}.  
In the constrained version of the \texttt{MSSM} (\texttt{mSUGRA}
boundary conditions) one finds only two candidates for the
\texttt{LSP}, namely the lightest neutralino or one of the right sleptons,
in particular the right scalar tau. Therefore we will consider below,
in detail, these two cases together with the possibility that a light
scalar top can have sizeable R-parity violating decays. However, if we
depart from the \texttt{mSUGRA} scenario, it has been shown
recently\cite{hirsch:2003fe} that the \texttt{LSP} can be of other
type, like a squark, gluino, chargino or even a scalar neutrino. In
section \ref{sec:otherLSP} we will briefly review their results.

\subsection{Probing Neutrino Mixing via Neutralino Decays}
\label{sec:neutralino-decays}

If R-parity is broken, the neutralino is unstable and it will decay
through the following channels: $\chiz{1} \to \nu_i \, \nu_j \,
\nu_k,\ \nu_i \, q \, \bar{q}, \ \nu_i \, l^+_j \, l^-_k, \ l^\pm_i \,
q \, \bar{q}', \ \nu_i \, \gamma $. It was shown\footnote{The relation
of the neutrino parameters to the decays of the neutralino has also
been considered in Ref.~\cite{Mukhopadhyaya:1998xj}.} in
Ref.~\cite{porod:2000hv}, that the neutralino decays well inside the
detectors and that the visible decay channels are quite large. This is
shown in \Fig{fig:neutra-decays}, where in the left panel we have the
neutralino decay length and on the right panel its invisible branching
ratio. We see that the invisible branching ratio stays always below
10\%, allowing most of the channels to be seen.
\FIGURE{
    \begin{tabular}{cc}
      \includegraphics[width=0.42\linewidth]{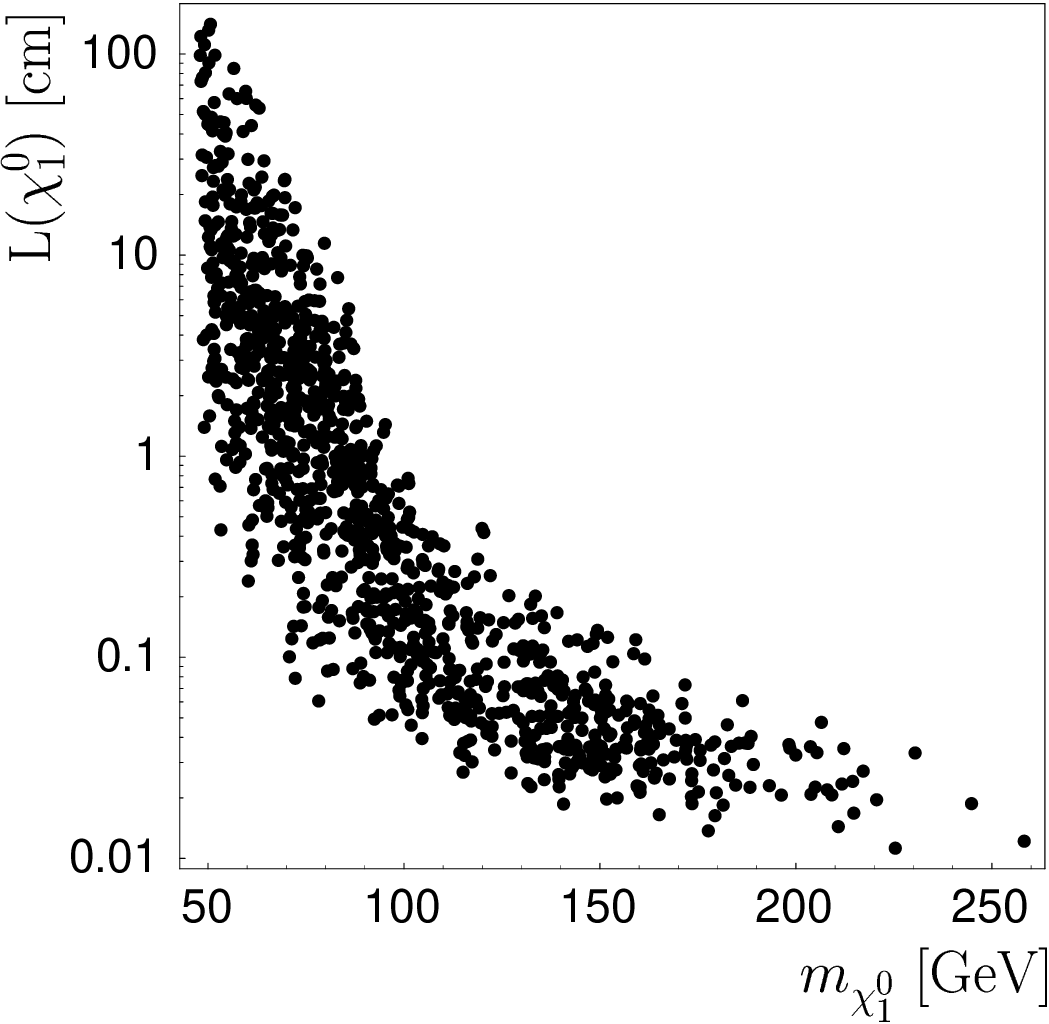}
      &
      \includegraphics[width=0.45\linewidth]{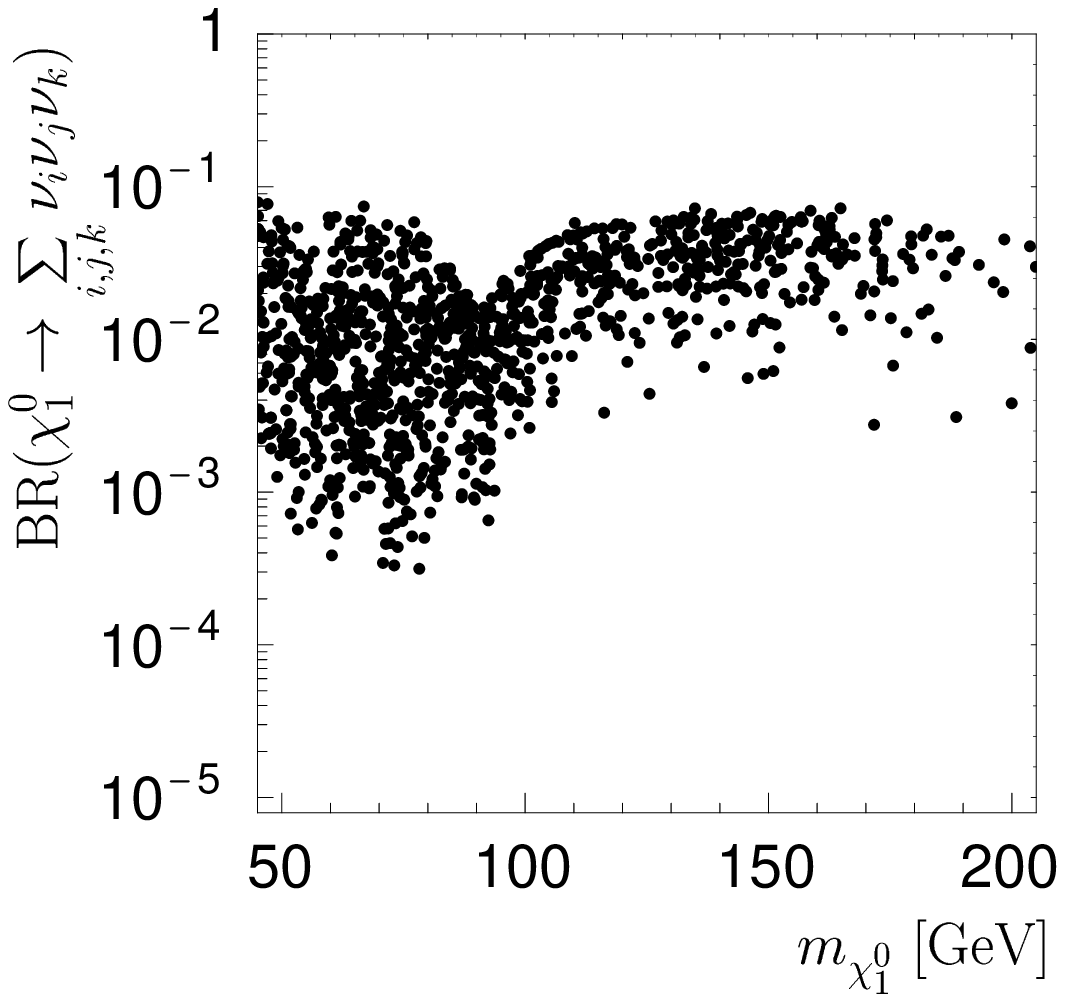}
    \end{tabular}
\caption{In the left panel we show the neutralino decay length. In the
      right panel it is shown the invisible neutralino branching ratio.}
\label{fig:neutra-decays}
}
We have shown in the previous sections that the ratios
$|\Lambda_i/\Lambda_j|$ and $|\epsilon_i/\epsilon_j|$ were very
important in the choice of solutions for the neutrino mixing
angles. What is exciting now, is that these ratios can be measured in
accelerator experiments.  In the left panel of Fig.~\ref{faro_fig3} we
show the ratio of branching ratios for semileptonic neutralino decays
into muons and taus: $BR(\chi \to \mu q' \bar q)/ BR(\chi \to \tau q'
\bar q$) as function of $\tan^2 \theta_{\Atm}$. We can see that there
is a strong correlation.  The spread in this figure can in fact be
explained by the fact that we do not know the \texttt{SUSY} parameters
and are scanning over the allowed parameter space. This is illustrated
in the right panel where we considered that \texttt{SUSY} was already
discovered with the following values for the parameters,
\begin{equation}
M_2=120\, GeV, \mu=500\, GeV, \tan\beta=5,
 m_0=500\, GeV, A=-500\, GeV
\label{eq:SUSYPOINT}
\end{equation}
\FIGURE{
\begin{tabular}{cc}
\includegraphics[width=0.45\textwidth]{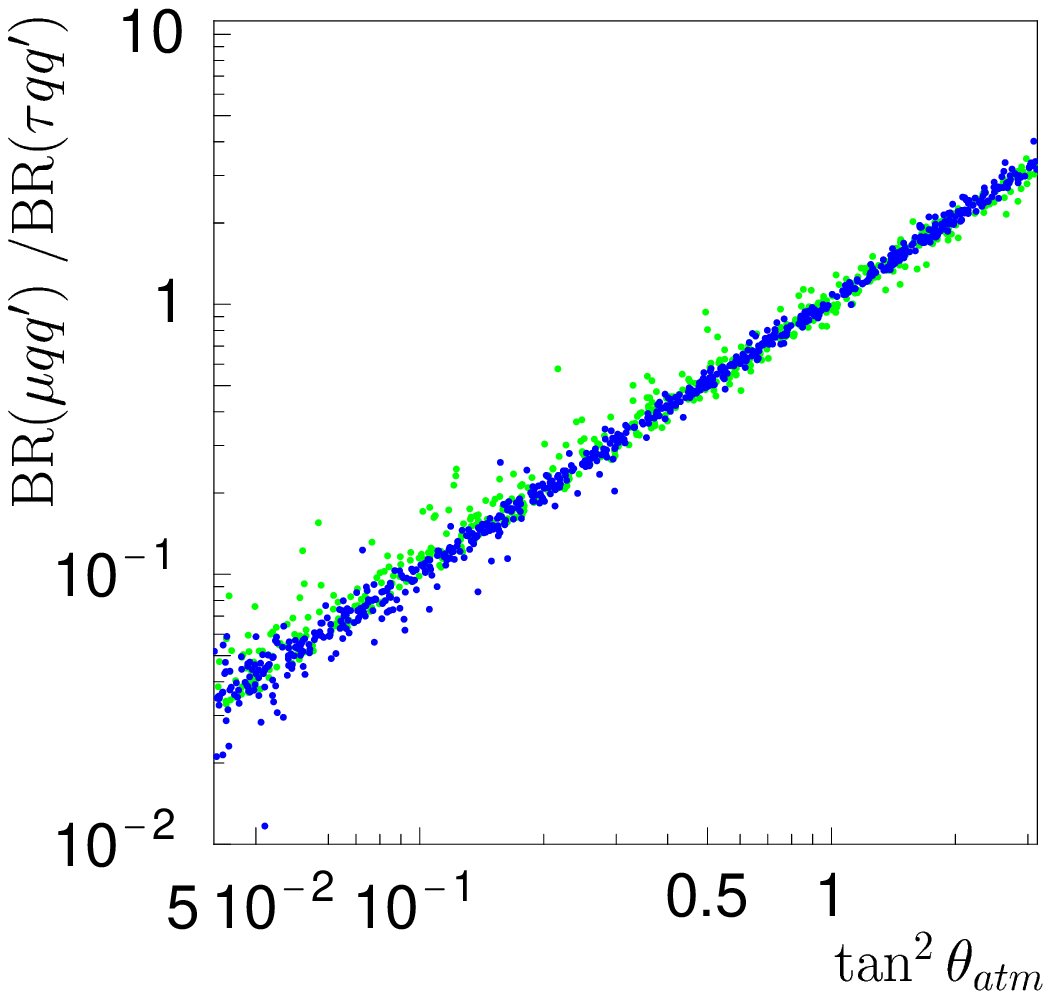}
&
\hskip -1mm
\includegraphics[width=0.45\textwidth]{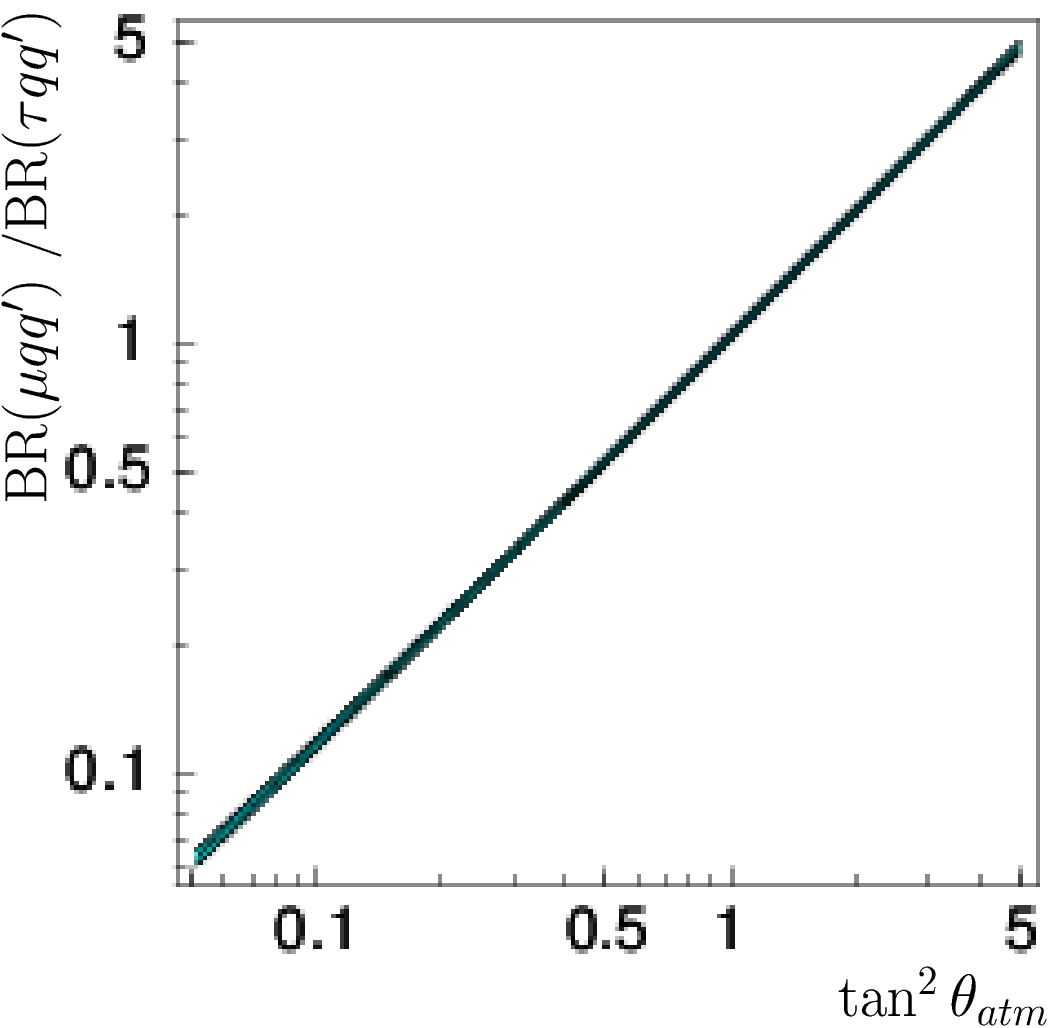}
\end{tabular}
\vspace{-3mm}
\caption{\small
Ratios of semileptonic branching ratios as a function of $\tan^2
\theta_{\Atm}$. On the left for random \texttt{SUSY} values and on the
right for the \texttt{SUSY} point of \Eq{eq:SUSYPOINT}.
}
\label{faro_fig3}
}
We see that the correlation is now extremely good and a measure of
those semileptonic branching ratios will be an important test for the
model.
\FIGURE{
\begin{tabular}{cc}
\includegraphics[width=0.45\textwidth]{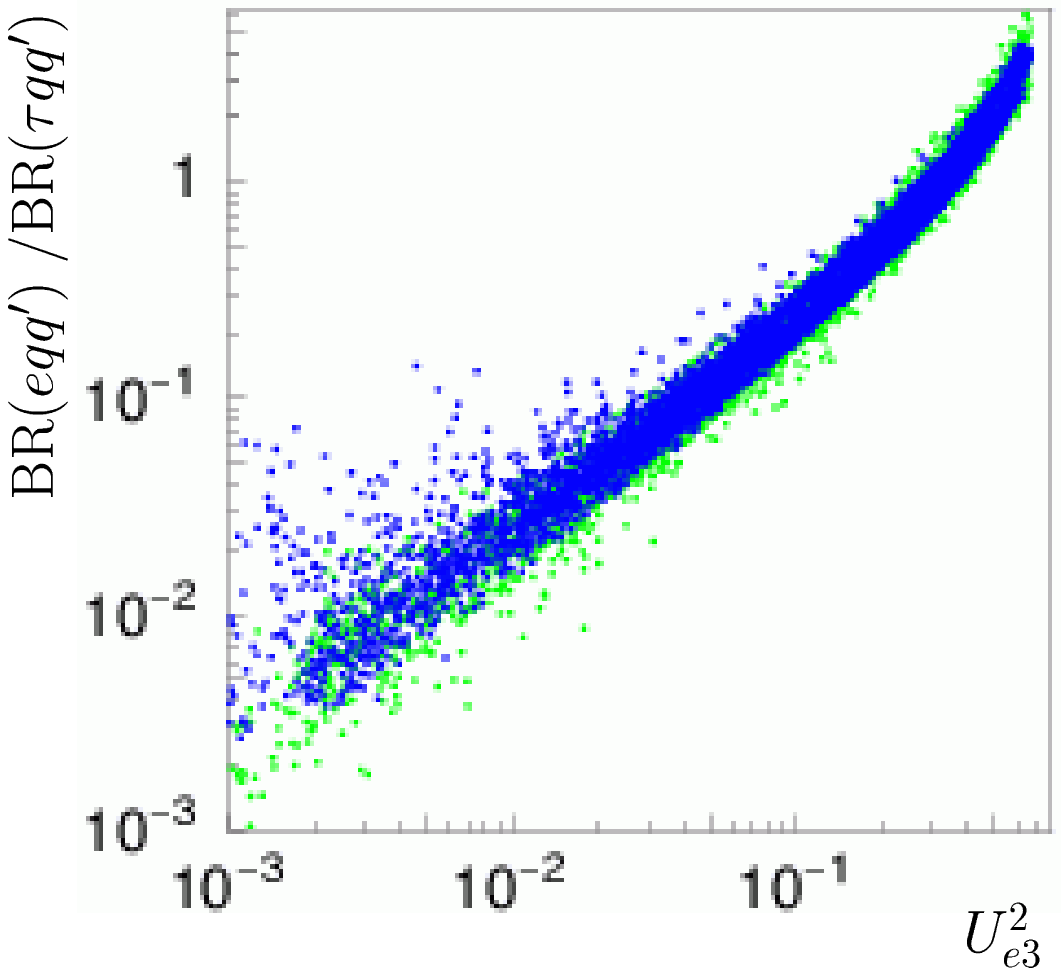}
&
\includegraphics[width=0.45\textwidth]{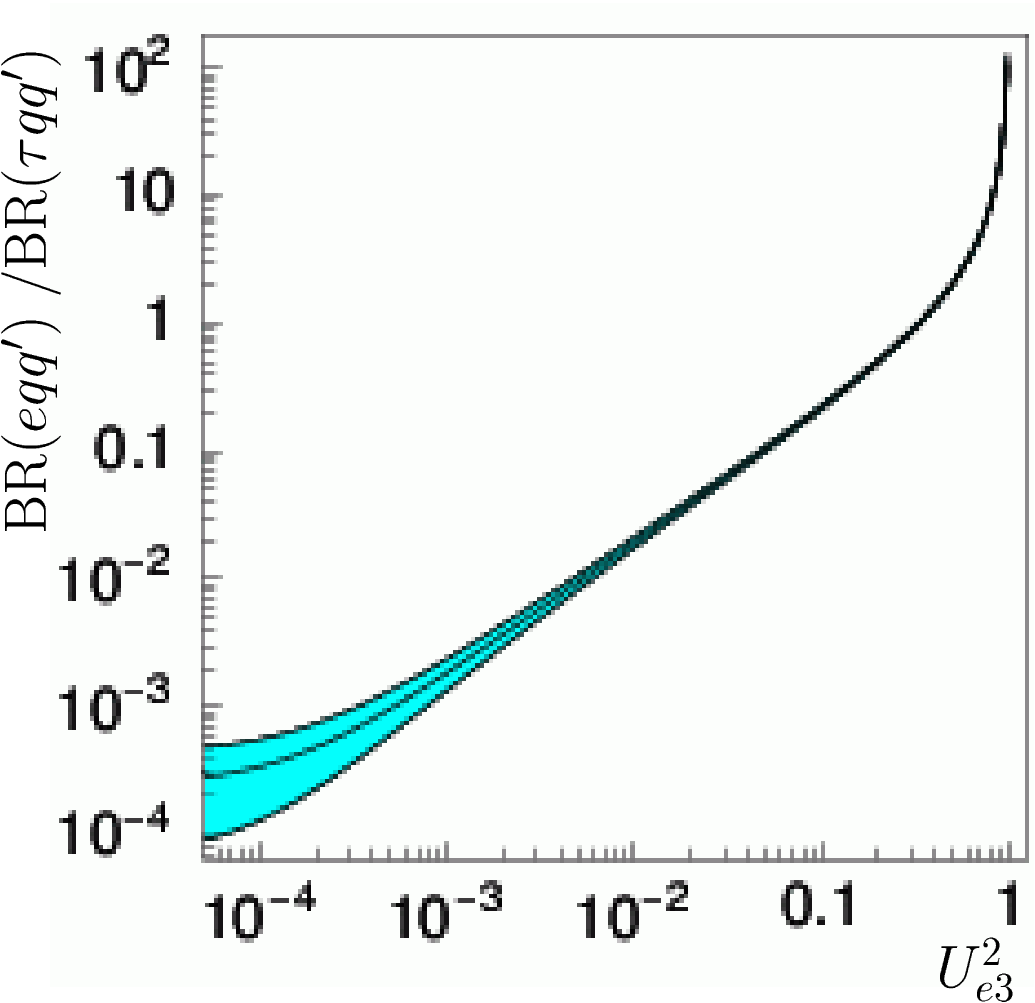}\\
\includegraphics[width=0.45\textwidth]{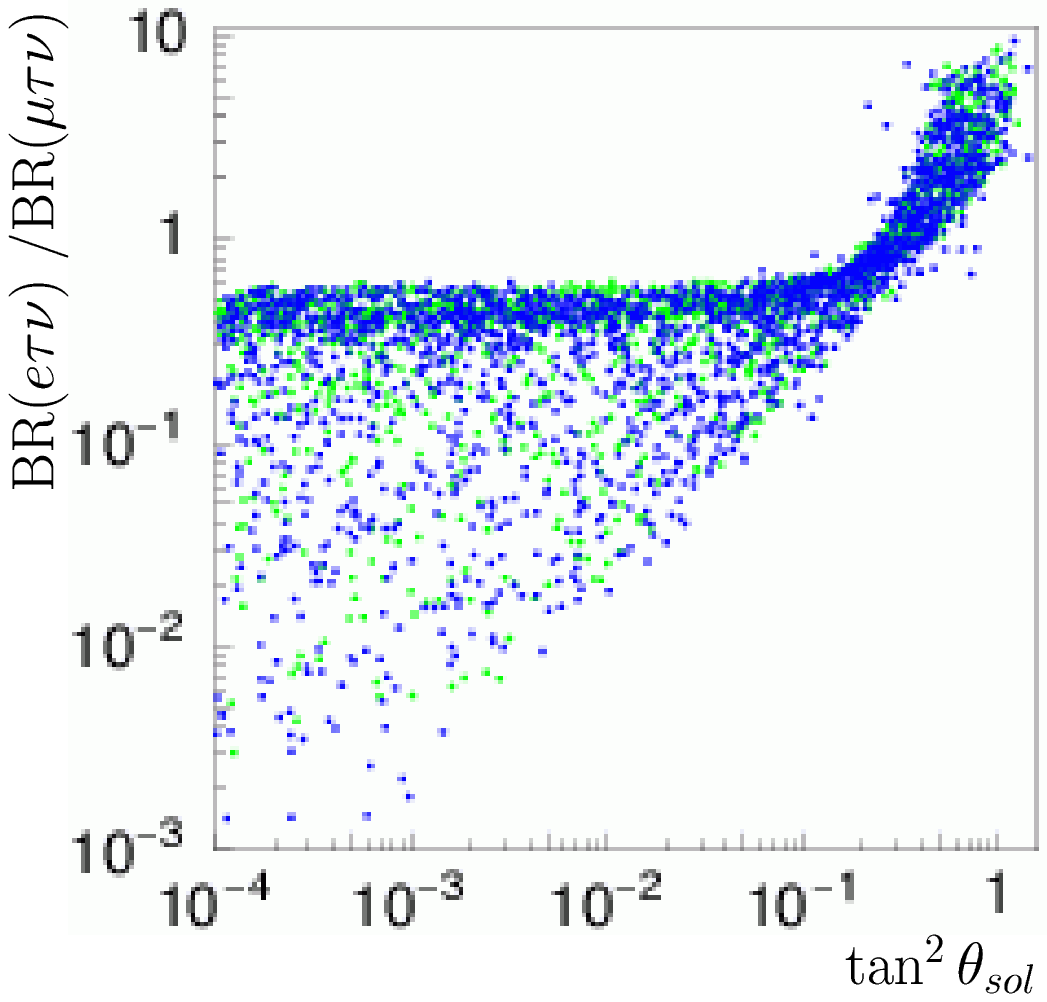}
&
\includegraphics[width=0.45\textwidth]{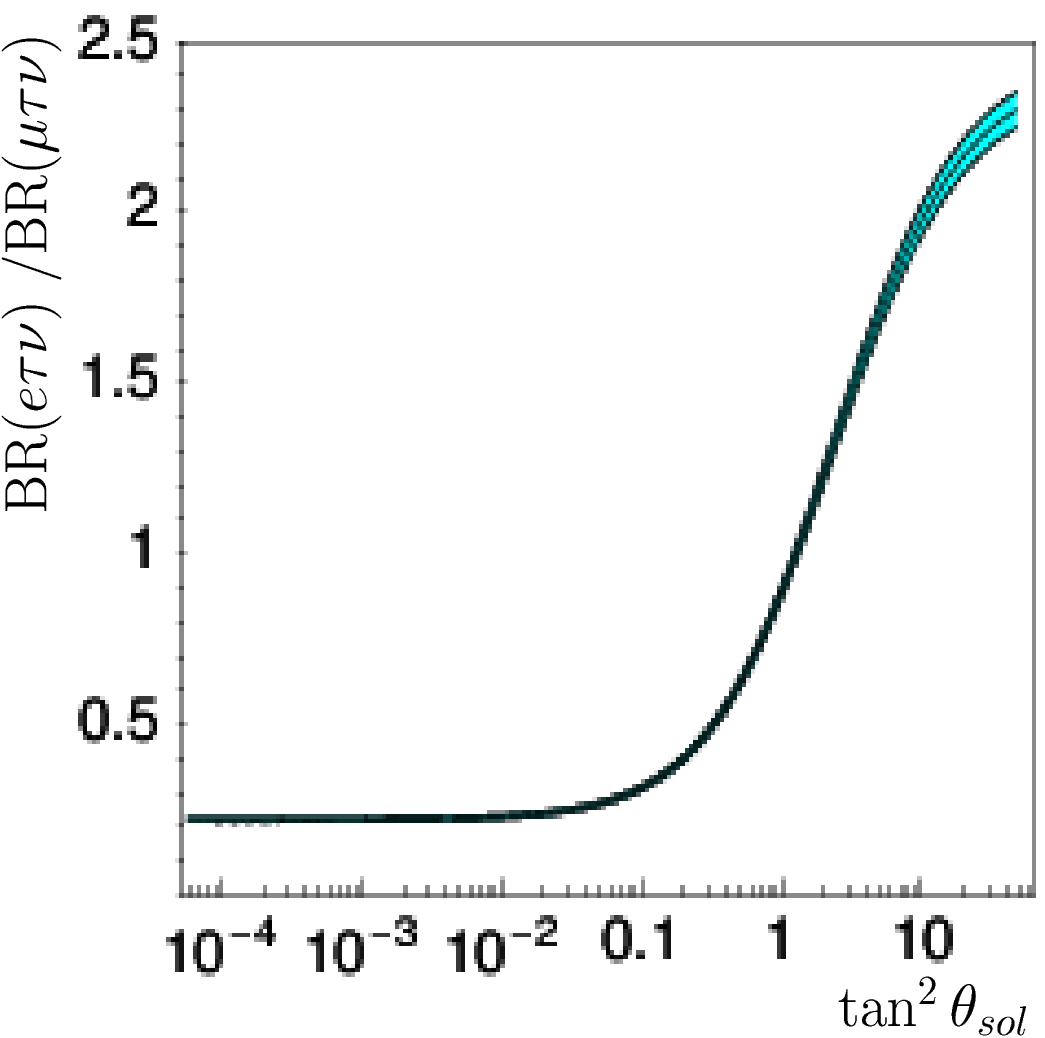}
\end{tabular}
\caption{Correlation of neutralino decays with $U^2_{e3}$ (top row)
  and with the solar angle (bottom row). On the left column we made a
  random scan over the \texttt{SUSY} parameter space, while on the right the
  parameters from \Eq{eq:SUSYPOINT} were assumed.}
\label{fig:xx1}
}
By looking at other decays we can look at other neutrino parameters,
as is indicated in \Fig{fig:xx1}. On the top row we have the
correlation with $U^2_{e3}$ and on the bottom row the correlation with
the solar angle. As before, the left panels correspond to a random
scan over the \texttt{SUSY} parameter space, while on the right panels we have
assumed that \texttt{SUSY} was already discovered. As an example we took the
\texttt{SUSY} parameters in \Eq{eq:SUSYPOINT}.

\subsection{Probing Neutrino Mixing via Charged Lepton Decays}

After considering the case of the \texttt{LSP} being the neutralino,
for completeness, we have also studied~\cite{hirsch:2002ys} the case
where a charged scalar lepton, most probably the scalar tau, is the
\texttt{LSP}.  We have considered the production and decays of $\st$,
$\te$ and $\tm$, and have shown that also for the case of charged
sleptons as \texttt{LSP}s they will decay well inside the
detector. This is shown in \Fig{faro_fig4}, where on the left panel we
show the production cross sections and on the right panel the decay
lengths for all the sleptons.
\FIGURE{
\begin{tabular}{cc}
\includegraphics[width=0.45\textwidth]{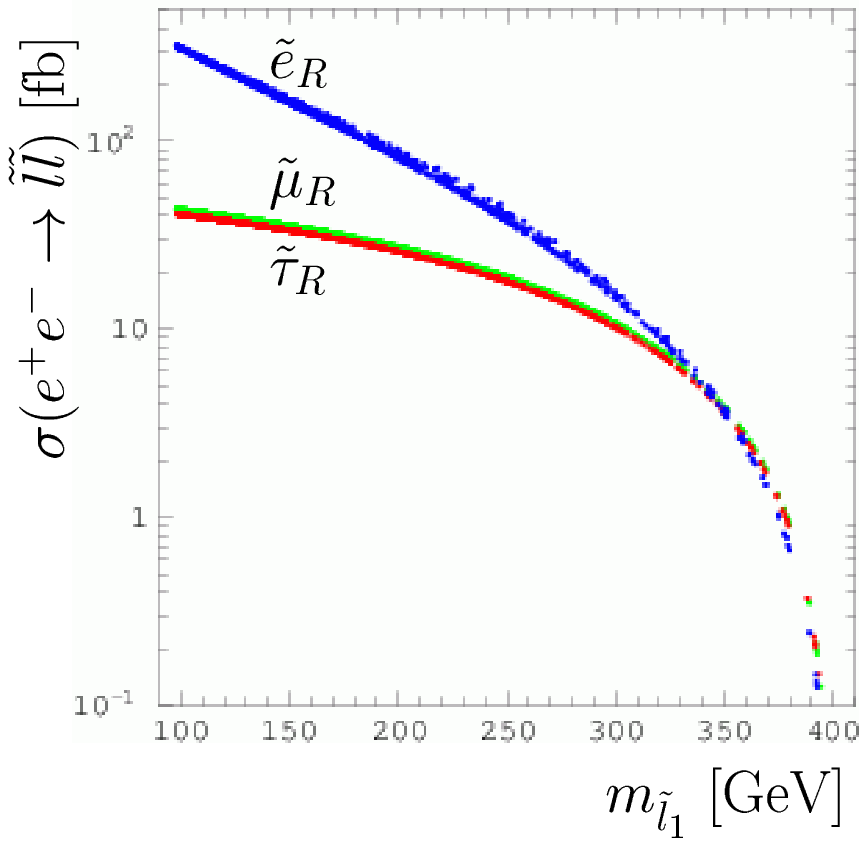}
&
\hskip -3mm
\includegraphics[width=0.45\textwidth]{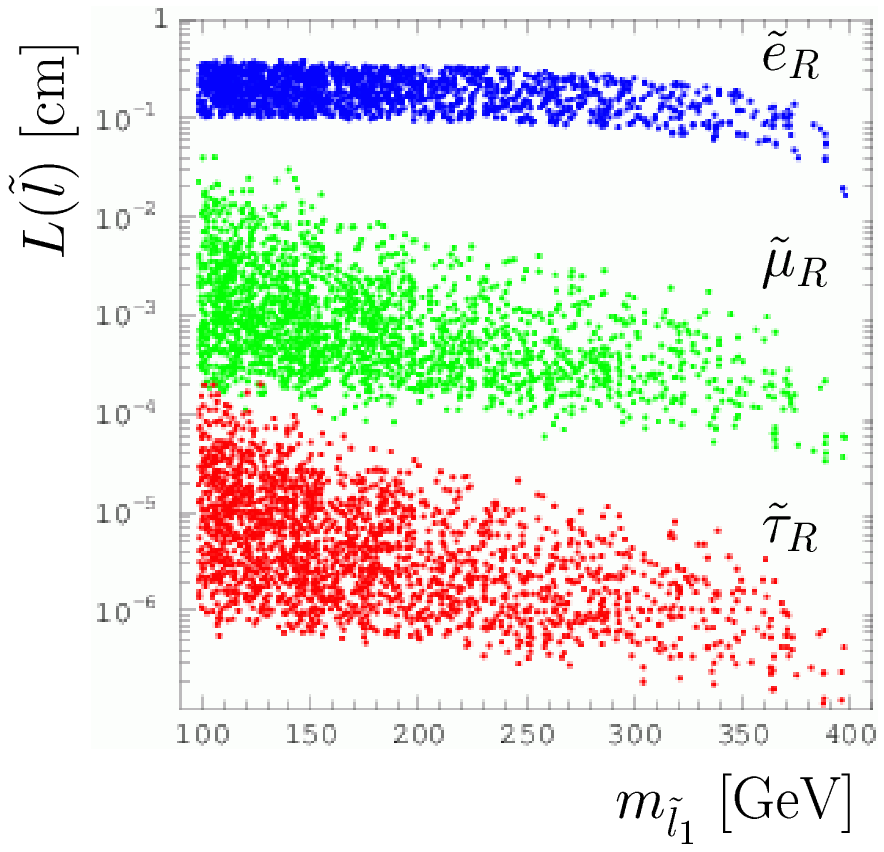}
\end{tabular}
\vspace{-3mm}
\caption{\small
a)$e^+ e^- \to {\tilde l}{\tilde l}$ production 
  cross section  at a Linear Collider $\sqrt{s}= 0.8 \hbox{TeV}$,
b) Charged slepton decay length at a linear collider with $\sqrt{s}=
  0.8 \hbox{TeV}$.
}
\label{faro_fig4}
}
After showing that the sleptons will decay inside the detector we can
correlate branching ratios with neutrino properties.  This is shown in
\Fig{faro_fig5} where the branching ratios of the scalar taus are
correlated with the solar angle, showing a strong correlation.

\FIGURE{
\begin{tabular}{cc}
\includegraphics[width=0.45\textwidth,height=55mm]{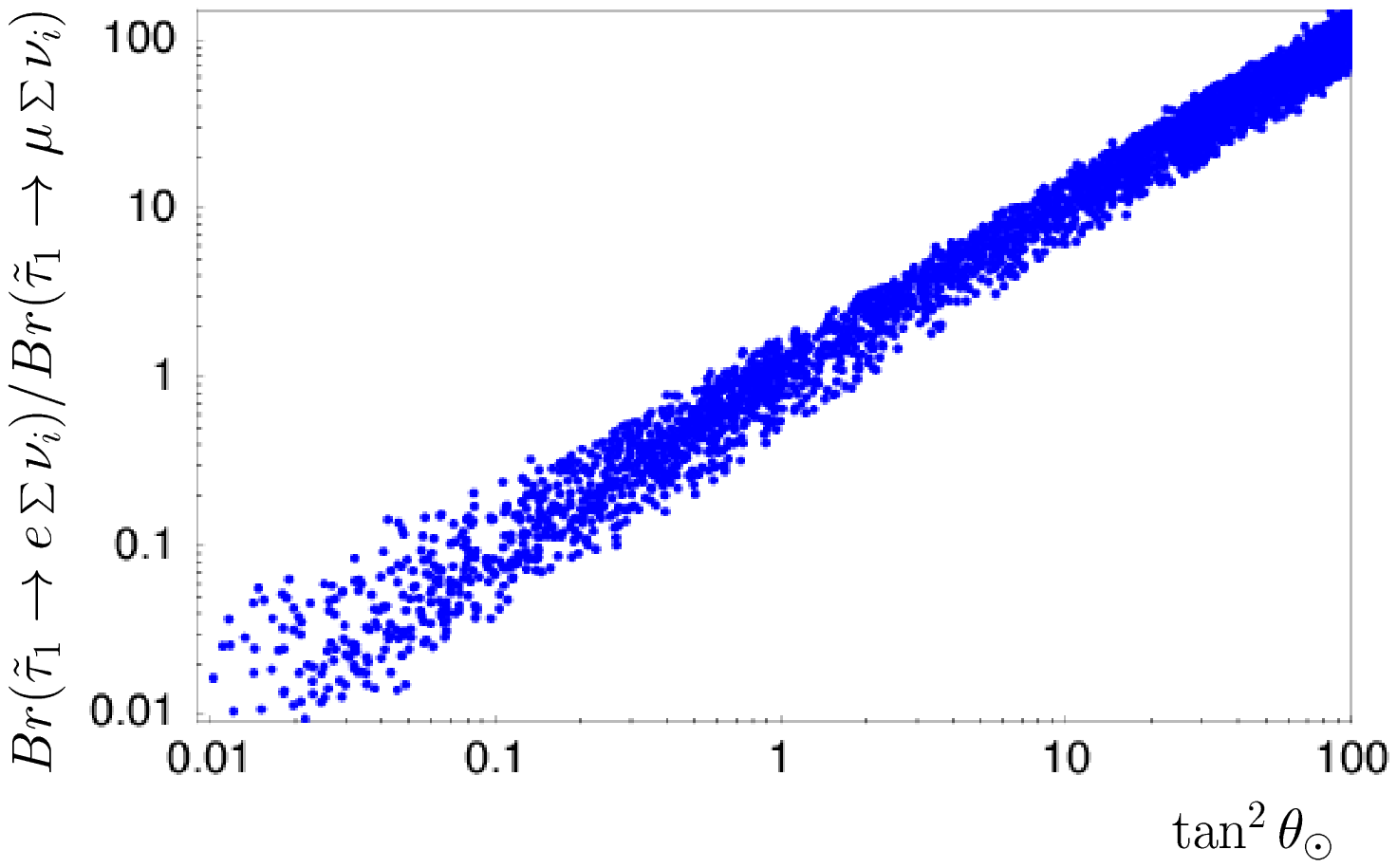}
&
\hskip -3mm
\includegraphics[width=0.45\textwidth,height=55mm]{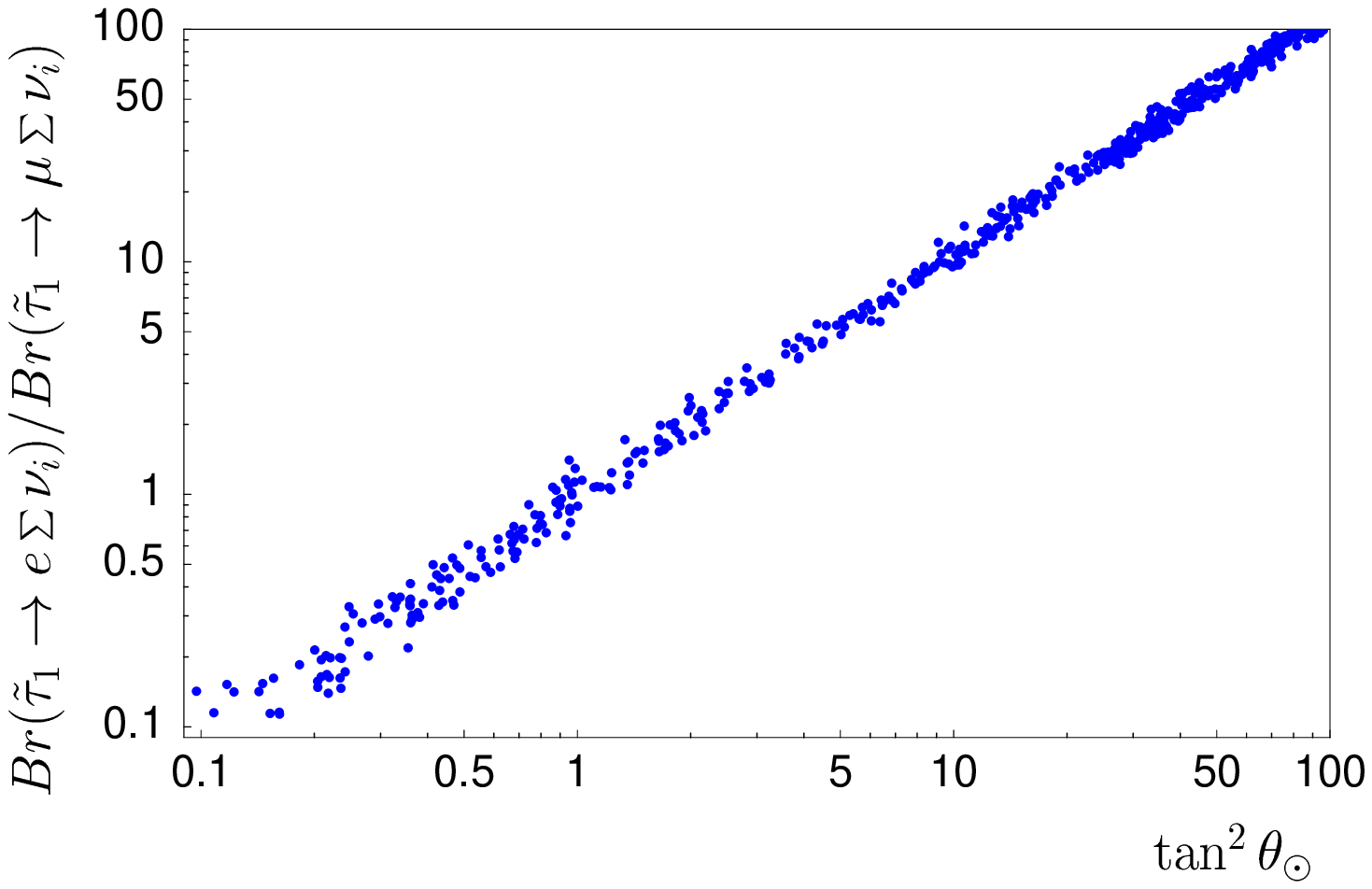}
\end{tabular}
\vspace{-3mm}
\caption{\small
Ratios of branching ratios for scalar tau decays 
  versus $\tan^2\theta_{\odot}$. The left panel shown all data
  points, the right one refers only to data points with
  $\epsilon_2/\epsilon_3$ restricted to the range [0.9,1.1].
}
\label{faro_fig5}
}

\subsection{Stop Decays and the Solar Angle}

It was shown~\cite{restrepo:2001me} that the stop decays are
complementary to $\chi^0$ decays as probes of the neutrino properties.
The semileptonic decays of the stop can be important as it is shown in
the left panel of Fig.~\ref{fig:stops}, taken from
Ref.~\cite{restrepo:2001me}, where it is plotted the $\tan \beta$
dependence of the branching ratio for the decay of $\tilde t_1$ into
$b\tau^+$ for several values of the neutrino mass. For
$m_{\nu_3}=0.06\,$eV, the $B(\tilde t_1\to b\,\tau)$ is still above
$0.1\%$ if $\tan\beta$ is not too large.  In the right panel of
Fig.~\ref{fig:stops} we show the ratio of B$({\tilde t}_1 \to b \,
e^+)/$B$({\tilde t}_1 \to b \, \mu^+)$ versus
$(\epsilon_1/\epsilon_2)^2$ for different values of $\cos
\theta_{\tilde t}$. For definiteness we have fixed the heaviest
neutrino mass at the best-fit value indicated by the atmospheric
neutrino anomaly.
One can see that the dependence is nearly linear even for rather small
$\cos \theta_{\tilde t}$.  One sees from the figure that, as long as
$\cos \theta_{\tilde t} \gtrsim 10^{-2}$ there is a good degree of
correlation between the branching ratios into $B(\tilde t_1 \to
b\,e^+)$ and $B(\tilde t_1 \to b\,\mu^+)$ and the ratio
$(\epsilon_1/\epsilon_2)^2$. Thus by measuring these branchings one
will get information on the solar neutrino mixing, since $\tan^2
\theta_{\Sol}$ is proportional to
$(\epsilon_1/\epsilon_2)^2$~\cite{romao:1999up}.  As a result in this
model one can directly test the solution of the solar neutrino problem
against the lighter stop decay pattern.
 \FIGURE{
   \begin{tabular}{cc}
     \includegraphics[clip,height=70mm]{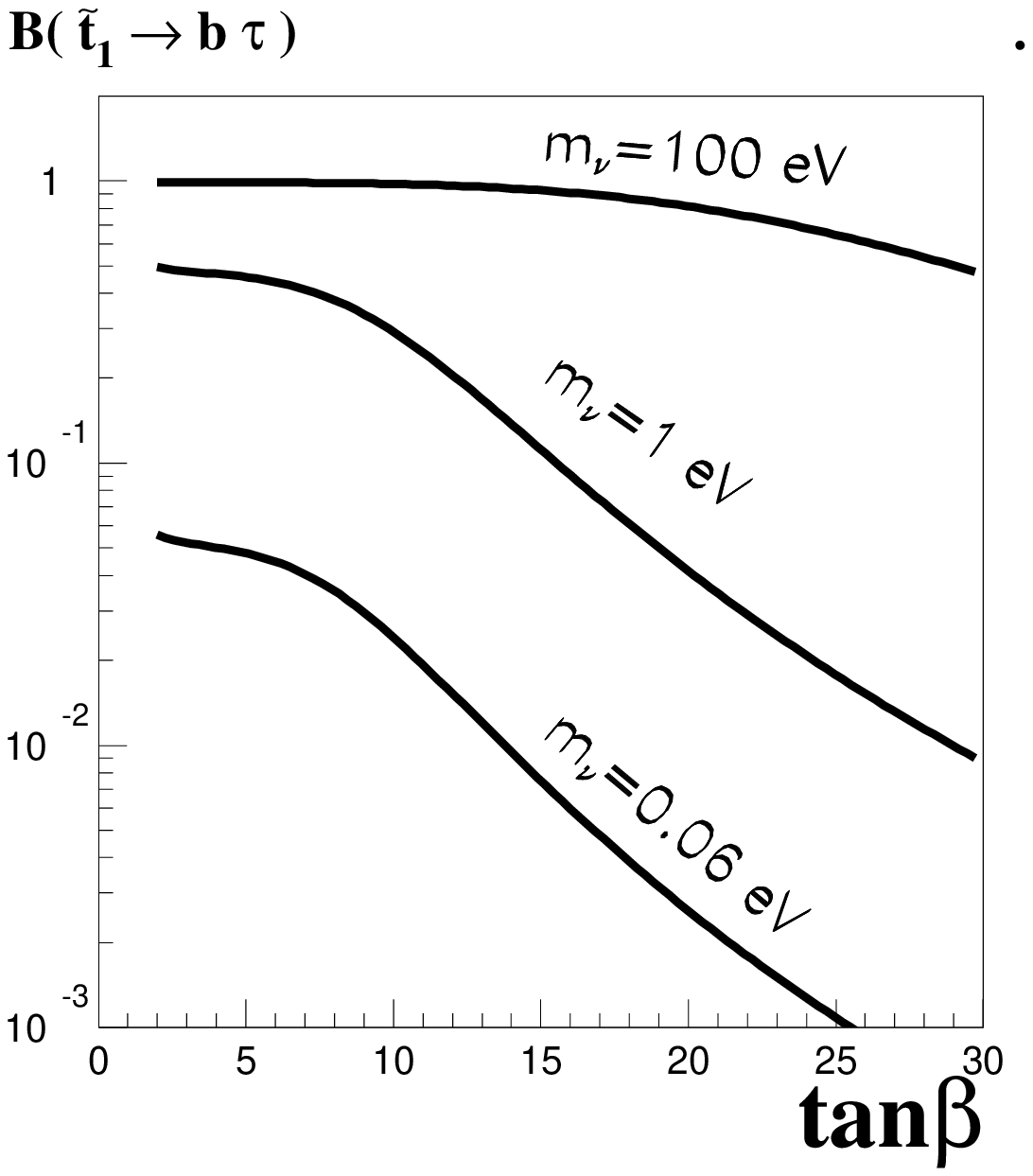}
     &
     \includegraphics[clip,height=65mm,width=65mm]{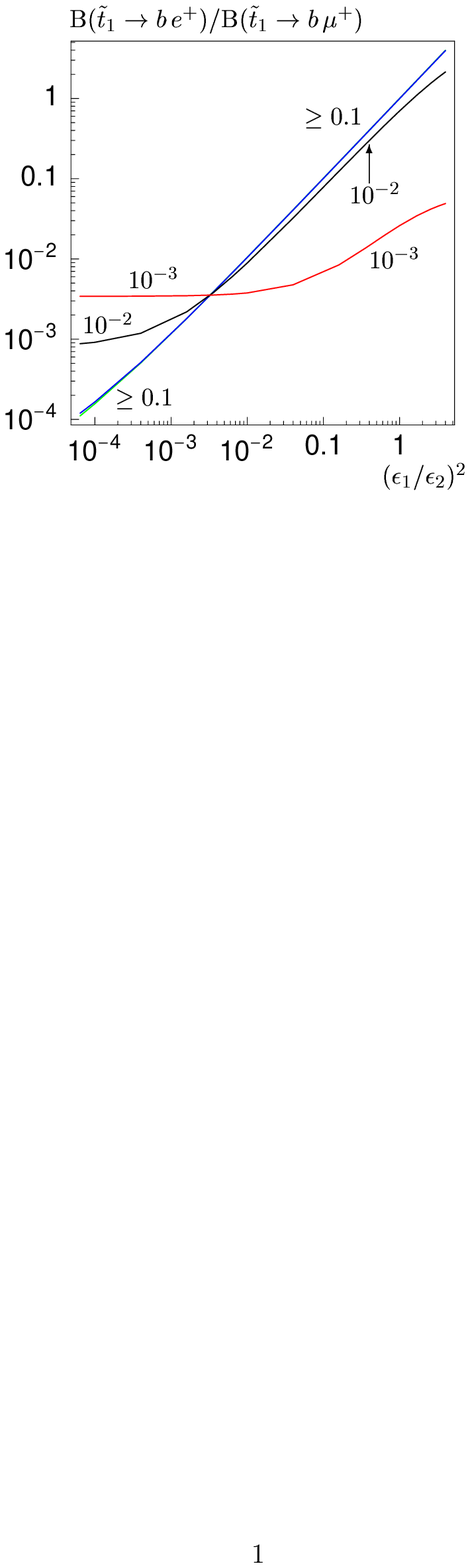}
   \end{tabular}
 \caption{On the left panel we plot the 
   branching ratios for ${\tilde t}_1$ decays
   for $m_{{\tilde t}_1} = 220$~GeV, $\mu = 500$~GeV, $M =
   240$~GeV, and $m_\nu = 100$, 1 and $0.06\,$eV for
   $\cos\theta_{\tilde t}=-0.8$. 
   On the right panel we plot the ratios: $\mbox{B}({\tilde t}_1 \to b e^+)/
   \mbox{B}({\tilde t}_1 \to b \mu^+) $ as a function of $(\epsilon_1 /
   \epsilon_2)^2$ for $m_{\nu_3} = 0.6$~eV and 
   $|\cos \theta_{\tilde t}| \geq 0.1, 0.01, 10^{-3}$. These plots
   were taken from Ref.~\cite{restrepo:2001me}}
 \label{fig:stops}
 }
This is also complementary to the case of neutralino decays considered
in section~\ref{sec:neutralino-decays}.  In that case the sensitivity
is mainly to atmospheric mixing, as opposed to solar mixing as can be
seen from Figs.~\ref{faro_fig3} and \ref{fig:xx1}. Testing the solar
mixing in neutralino decays at a collider experiment requires more
detailed information on the complete spectrum to test the solar angle
\cite{porod:2000hv}. In contrast we have obtained here a rather neat
connection of stop decays with the solar neutrino physics.

\subsection{Other \texttt{LSP} decays}
\label{sec:otherLSP}

As we discussed in the introduction to this section, if we depart from
the \texttt{mSUGRA} scenario, then the \texttt{LSP} can be of other type,
like a squark, gluino, chargino or even a scalar neutrino, as it has
been shown recently in Ref.~\cite{hirsch:2003fe}. As the decays of
these \texttt{LSP}'s will always depend on the parameters that
violate R-parity and induce neutrino masses and mixings, it is
possible to correlate branching ratios to the neutrino
properties. This is shown in \Fig{fig:martin-werner} taken from
Ref~\cite{hirsch:2003fe}. On the left panel we consider the case of
chargino decays and show $BR(\tilde\chi^+ \rightarrow \mu \bar{c} c)/
BR(\tilde\chi^+ \rightarrow \tau \bar{c} c)$ as a function of the
ratio $(\Lambda_2/\Lambda_3)^2$. As we saw in
section~\ref{sec:neutrino-results}, this last ratio is correlated with
the atmospheric angle. So, by looking at the chargino decays one can
test the atmospheric mixing angle. In fact, as the authors of
Ref~\cite{hirsch:2003fe} have shown, one can use the already very
precise data on the atmospheric mixing angle to put bounds on the ratios of
several branching ratios of chargino decays. On the right panel we
show a very strong correlation for the solar mixing angle
obtained\cite{hirsch:2003fe} with the decays of squarks.
Many other correlations can be obtained making the model over constrained.
So, in summary, no matter what supersymmetric particle is the
\texttt{LSP}, measurements of branching ratios at future accelerators
will provide a definite test of the \texttt{BRpV} model as a viable model
for explaining the neutrino properties.

\FIGURE{
\begin{tabular}{cc}
\includegraphics[width=0.45\textwidth,height=60mm]{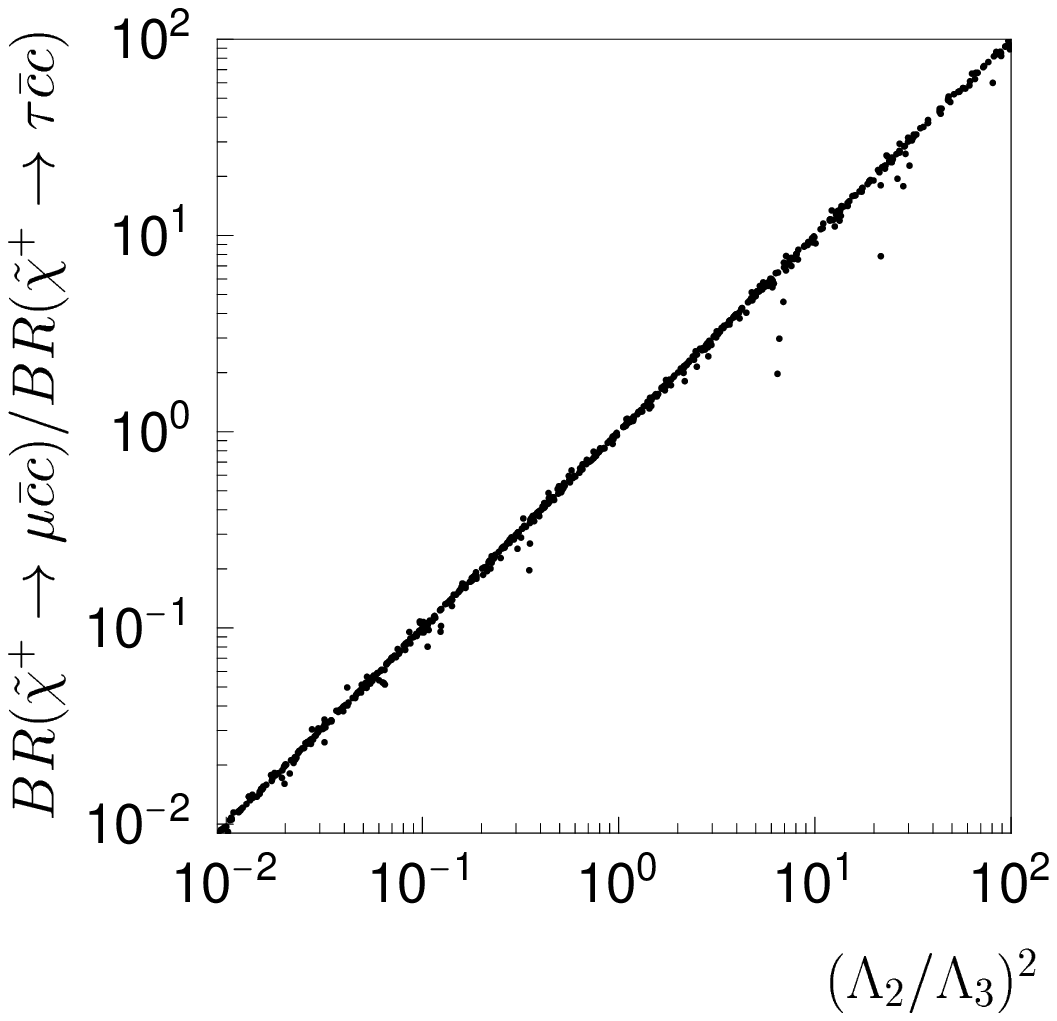}
&
\hskip -3mm
\includegraphics[width=0.45\textwidth,height=60mm]{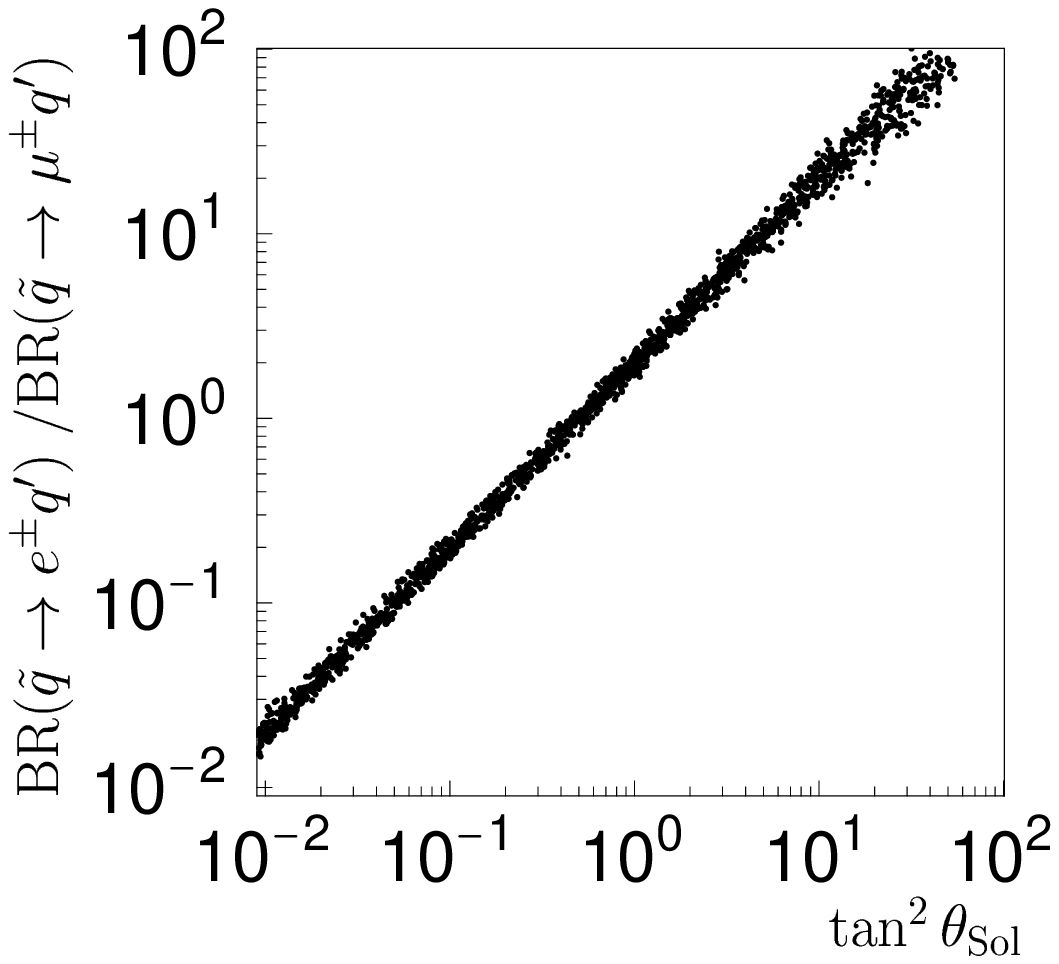}
\end{tabular}
\vspace{-3mm}
\caption{\small
Left panel: ratios of branching ratios for chargino decays 
  versus $(\Lambda_2/\Lambda_3)^2$; Right panel:
ratios of branching ratios for squark decays 
  versus $\tan^2\theta\Sol$. These plots were taken from
  Ref.~\cite{hirsch:2003fe}. }
\label{fig:martin-werner}
}

\section{Conclusions}

The Bilinear R-Parity Violation Model is a simple extension of the
\texttt{MSSM} that leads to a very rich phenomenology.  Hopefully, it
will be an effective model for the more theoretically attractive case
where R-parity is spontaneously broken~\cite{romao:1992vu,romao:1992ex}.

We have calculated the one--loop corrected masses and mixings for the
neutrinos in a completely consistent way, including the RG equations
and correctly minimizing the potential.  We have shown that it is
possible to get easily maximal mixing for the atmospheric neutrinos
and large angle MSW, as it is preferred by the present neutrino
data. We have also obtained approximate formulas for the solar mass
and solar mixing angle, that we found to be very good, precisely in the
region of parameters favored by this data.

We emphasize that the \texttt{LSP} decays inside the detectors, thus
leading to a very different phenomenology than the \texttt{MSSM}. In
the \texttt{mSUGRA} scenarios, the \texttt{LSP} can be either the
lightest neutralino, like in the \texttt{MSSM}, or a charged particle,
must probably the lightest stau. In both cases we have shown that
ratios of the branching ratios of the \texttt{LSP} can be correlated
with the neutrino parameters.  We have also shown, that the decays of
the stop can be complementary to those of the \texttt{LSP} for the
case of the solar mixing angle.

In more general scenarios than the constrained \texttt{mSUGRA}, the
\texttt{LSP} can essentially be any supersymmetric particle. However,
also in these scenarios the neutrino properties can be correlated with
ratios of branching ratios for these particles.

So, in summary, no matter what supersymmetric particle is the
\texttt{LSP} measurements of branching ratios at future accelerators
will provide a definite test of the \texttt{BRpV} model as a viable model
for explaining the neutrino properties.

\acknowledgments

This work was partially supported by the
European Commission RTN grant HPRN-CT-2000-00148.


\end{document}